\documentclass[12pt,preprint]{aastex}
\def \lp{\>\> .}
\def \lc{\>\> ,}

\def \c2{cm$^{-2}$}
\def \cc{cm$^{-3}$}
\def \kms{kms$^{-1}$}
\def \s{s$^{-1}$}

\def \nh3{NH$_3$}
\def \n2h{N$_2$H$^+$}

\def \nh2{n_{H_2}}
\def \nh1{n_{HI}}

\def \tw{$^{12}$CO}
\def \th{$^{13}$CO}
\def \ce{C$^{18}$O}
\def \h2{H$_2$}

\def \Ms{$M_{\odot}$}

\def \be{\begin{equation}}
\def \ee{\end{equation}}
\def \bf{\begin{figure}}
\def \ef{\end{figure}}

\def \lc{\>\> ,}
\def \lp{\>\> .}

\begin{document}

%

\shorttitle{}
\shortauthors{Goldsmith, Li, \& Krco}

\title{The Transition from Atomic to Molecular Hydrogen in Interstellar Clouds:  21cm Signature of the Evolution of Cold Atomic Hydrogen in Dense Clouds}

\author{Paul F. Goldsmith\altaffilmark{1}, Di Li\altaffilmark{2}, and 
Marko Kr\v{c}o\altaffilmark{3}}
\altaffiltext{1}{JPL, 4800 Oak Grove Drive, Pasadena CA 91109, Paul.F.Goldsmith@jpl.nasa.gov}
\altaffiltext{2}{JPL, 4800 Oak Grove Drive, Pasadena CA 91109, dli@jpl.nasa.gov}
\altaffiltext{3}{Department of Astronomy, Cornell University, Ithaca NY 14853 marko@astro.cornell.edu }

\begin{abstract}

We have investigated the time scale for formation of molecular clouds by examining the conversion of HI to \h2\ using
a time--dependent model which includes \h2\ photodissociation with rate dependent on dust extinction and self shielding. 
\h2 formation on dust grains and cosmic ray destruction are also included in one--dimensional model slab clouds which incorporate time--independent density and temperature distributions.
We calculate 21cm spectral line profiles seen in absorption against a background provided by general Galactic HI emission, and compare the model spectra with HI Narrow Self--Absorption, or HINSA, profiles absorbed in a number of nearby molecular clouds.
The time evolution of the HI and \h2\ densities is dramatic, with the atomic hydrogen disappearing in a wave propagating from the central, denser regions which have a shorter \h2\ formation time scale, to the edges, where the density is lower and the time scale for \h2\ formation longer.
The model 21cm spectra are characterized by very strong absorption at early times, when the HI column density through the model clouds is extremely large.
Excess emission produced by the warm edges of the cloud when the background temperature is relatively low can be highly confusing in terms of separating the effect of the foreground cloud from variations in the background spectrum.  
The minimum time required for a cloud to have evolved to its observed configuration, based on the model spectra, is set by the requirement that most of the HI in the outer portions of the cloud, which otherwise overwhelms the narrow absorption, be removed.
The characteristic time that has elapsed since cloud compression and initiation of the HI $\rightarrow$ \h2\ conversion is a few $\times$ 10$^{14}$ s or $\simeq$ $10^7$ yr.
This sets a minimum time for the age of these molecular clouds and thus for star formation that may take place within them.

\end{abstract}

\setcounter{footnote}{0}

\section{INTRODUCTION}

It is generally accepted that stars form from dense interstellar clouds
in which hydrogen and most other elements are largely in molecular form.  
Studies of the later phases of star formation have progressed rapidly thanks to improved
instrumental angular resolution, sensitivity, and frequency coverage, but the formation of 
molecular clouds themselves has been rather neglected of late, and this remains a major
poorly understood step in the overall process of converting interstellar material into young stars.  
A variety of processes have been postulated to explain molecular
cloud formation, many of which are reviewed by Elmegreen (1991).
The majority of the scenarios for forming molecular clouds (whether 
massive GMCs or less massive dark clouds) involve compressing and increasing the
column density of previously atomic material, thus reducing the photodestruction rate of 
\h2 (and other) molecules, and thus leading to the transformation from atomic to predominantly
molecular form.

It is of considerable interest to determine whether there is any signature in
molecular clouds of their past history as primarily atomic objects, and to examine
whether any information can be extracted about their evolutionary time scale.  
It was pointed out more than 30 years ago by Shu (1973), that the atomic hydrogen
in molecular clouds may be a residue of their previous state, and as such could
be used to place an upper limit on their age of $\simeq$ 10$^7$ yr.  

One powerful tool for studying atomic hydrogen is absorption of radiation from a distant source by relatively cool HI in a nearby cloud.  
If the background radiation is itself HI emission, the foreground cloud is said to be producing HI self absorption, or HISA.  
The literature on HI self absorption is currently so large that a reasonably
complete listing of papers would be very difficult and probably not immediately helpful.
In previous papers (Li \& Goldsmith 2003; hereafter Paper I and Goldsmith \& Li 2005; hereafter Paper II) we referenced a number of papers that have focused particularly on the issue of HI narrow self absorption, as this is directly relevant to the issue of the relationship of atomic and molecular clouds. 
Some early papers that were not referenced in Paper I or Paper II but should have been are those of \cite{heeschen1955}, \cite{radhakrishnan1960}, and \cite{sancisi1970}.
In addition to the references cited in Papers I and II, the relationship of atomic gas traced through HI absorption, and molecular gas traced by carbon monoxide emission was studied by \cite{sato1978}, \cite{liszt1981}, \cite{peters1987}, \cite{hasegawa1983}, \cite{garwood1989}, and \cite{feldt1993}. 

Observations of both the atomic and molecular constituents of dense clouds have
improved considerably in recent years (e.g. Paper I).
These authors found that dense condensations (or cores) within molecular
clouds contain significant amounts of atomic hydrogen, which in terms of line
velocity, line width, and line depth, could be identified as coming from cold ($T$ $\le$ 20 K) regions, which are well shielded from the interstellar radiation field.
The term HI Narrow Self Absorption, or HINSA, was used to describe this 
component of atomic hydrogen in the interstellar medium, with the narrow being
defined as having nonthermal line width no larger than that of accompanying \tw,
whose presence is required for HINSA.
In Figure 1 we show some examples of HINSA spectra in relatively nearby dark cloud cores.
Not included are the corresponding spectra of molecular emission. 
The velocities and nonthermal line widths of OH, \th, and \ce\ agree
very closely with those determined for HINSA features.
\clearpage
\thispagestyle{empty}
\bf
\vspace*{-15mm}
\plotone{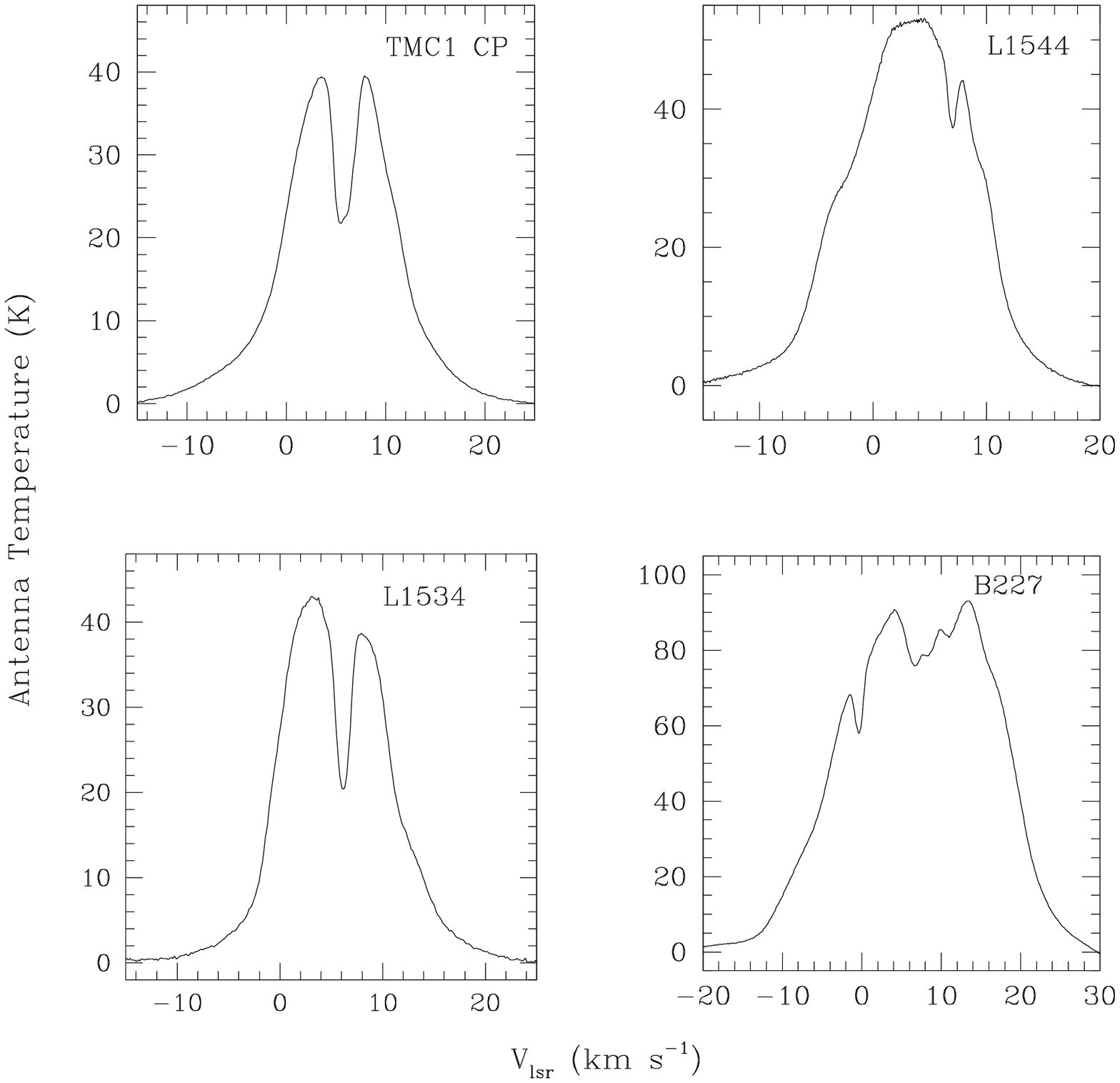}
\caption{\label{spectra} HI spectra of four nearby dark cloud cores taken with the Arecibo Observatory 305 m 
telescope.  The data have been corrected for the main beam efficiency of 0.60.  
The spectra are at the central position of each source, with coordinates given
in Papers I and II.  
The rms statistical noise in these spectra is approximately 0.09 K (5 min. on source integration with 0.382 kHz channel bandwidth) off the HI emission features, and 2 to 3 times this level on the lines.  
Thus is still less than 1 \% of the emission intensity, and thus 30 times smaller than even ``weak'' HINSA features.  
The statistical noise is barely visible on the flat--topped portion of the L1544 spectrum.  The cold HI coextensive with the molecular component of each cloud core produces the narrow absorption feature seen in each spectrum.  
The absorption profile in TMC1 CP comprises two (marginally resolved) velocity components which together significantly broaden the observed line profile.  
Various complex spectral structure with less rapid variation of intensity as a function of velocity is visible in some sources, including in the 4 to 14 \kms\ velocity range in B227. Its origin is much less certain than that of the sharp, narrow absorption, and may be a combination of variations in the temperature of the background emission gas and absorption by intervening, atomic gas with T $\leq$ 80 K.
}
\ef 
\clearpage

A subsequent paper (Goldsmith \& Li 2005) used detailed observations
of the distribution and quantity of cold HI within relatively well--isolated molecular 
cores to determine their ages to be between 3$\times$10$^6$ yr and $\ge$ 3$\times$10$^7$ yr.  
In this context, ``age'' refers to the time since the clouds, initially having hydrogen
completely in atomic form due to rapid destruction of molecular hydrogen by photodestruction, had this source of \h2 destruction instantaneously turned off, and were allowed to evolve with the hydrogen gradually becoming molecular.  
Destruction of \h2 by cosmic rays maintains a small residual amount of atomic hydrogen at long times after the beginning of this evolutionary process.

The clouds studied in Paper II have an atomic hydrogen abundance slightly exceeding that predicted to characterize steady state conditions.  
The ``age'' that is derived (being close to the characteristic time to achieve a steady state HI/\h2 ratio) is indicative of a relatively long lifetime for these objects, and thus has significant potential implications for understanding the overall process of molecular cloud and star formation.  
It is thus worthwhile to examine cloud evolution in more detail, trying to take into
account the critical physical processes that determine the rate of conversion of 
atomic to molecular hydrogen.  
This is a complex issue, particularly when one considers that the 
line width, temperature, density, and fractional abundance variations all play a
role in determining the observed line profiles.

In the present work we take one step towards the ultimate goal of developing a complete model for the evolution of 21cm line profiles in evolving molecular clouds by employing a simplified model in which we do not include any dynamical evolution, considering a cloud to have fixed temperature and density profiles which are functions only of the distance from the surface of the cloud.  
These profiles are chosen to mimic those obtained in other studies for a steady state
solution including chemistry and thermal balance.   
We model the cloud as a one--sided slab, taking into account \h2 destruction
by the interstellar radiation field, together with shielding by dust and the critical process of self--shielding.  
Cosmic ray destruction of \h2 occurs throughout the cloud, as does \h2 formation on
grain surfaces, parameterized in a conventional manner.
Thus, our results predict only approximately what may be happening early in the evolution of
the system, and are most relevant for interpreting the later stages of the atomic to 
molecular conversion process.

We describe our model for cloud structure and evolution in \S\ref{model}.
We discuss the relevant molecular and atomic processes and present results for the distribution
of atomic and molecular gas in two different cloud models in \S\ref{results},
In \S\ref{cloud_spectra} we present 21cm spectral line profiles for model clouds observed in absorption, and we discuss the implications of our results in \S\ref{conclusions}.

\section{CLOUD MODEL WITH TIME--DEPENDENT HI AND H$_2$ ABUNDANCES}
\label{model}

The transition from the atomic to molecular phase of the
ISM may have, as discussed above, a variety of triggers.  
While star formation occurs in molecular clouds located throughout the Milky Way,
the rate of star formation is observed to be enhanced in spiral arms (e.g. Siegar \& James 2002).
The common feature is typically a rapid compression with accompanying increase in
extinction as described by \cite{elmegreen1996}.  
This has led to a number of studies in which atomic clouds evolving towards a molecular
state have been tentatively identified (Minter et al. 2001; Gibson et al. 2005; Kavars et al. 2005).  
Given the general belief that molecular clouds are finite--lived structures, another 
important initiator of cloud compression and the process of atomic to molecular hydrogen conversion is shocks produced by galactic spiral density
waves, discussed by Shu (1972) and recently modeled by Dobbs, Bonnell, \& Pringle (2006)  
\footnote{An alternative view is that large molecular clouds are assembled from
preexisting small molecular clouds, as discussed by \cite{pringle2001}
}
.
Outflows from young stars and shocks from expanding HII regions and supernovae have also been put forth as the agents responsible for what is sometimes called triggered or stimulated star formation (Elmegreen 1992; Patel et al.\ 1998; Karr \& Martin 2003; Hosokawa \& Inutsuka 2006).
Cloud collisions may drive compressive shocks into atomic clouds which can result in conversion of material to molecular form (e.g. Bergin et al.\ 2004). 
Compression due to shocks from colliding flows has been investigated extensively in recent years, including
the studies of Hennebelle \& P\'{e}rault (1999, 2000), Audit \& Hennebelle (2005), and \cite{vazquez2006}.
Instabilities may then result in the formation of ``individual'' molecular clouds (Koyama \& Inutsuka 2000, 2002) but these processes may not always result in gravitationally bound regions.

While it would be desirable to follow the post shock evolution of an interstellar 
cloud including dynamics, chemistry, and thermal balance, we here restrict ourselves
to the highly idealized situation of a cloud which is assumed at $t$ = 0 to be characterized by its steady state temperature and density distributions, and follow only the evolution of atomic to molecular hydrogen under these conditions. 
Although this represents a significant approximation, in most models the cloud compression is relatively rapid 
\footnote{
In Bergin et al. (2004) the time for shock compression is $\leq$ 1 million years for shock velocities greater than 15 \kms, but approaches 10 million years for the lowest shock velocity considered, 10 \kms}
. 
The cooling time is the ratio of the thermal energy per unit volume to the cooling rate per unit volume, 
\be
\tau_{cool} = \frac{3/2 n k T}{\Lambda} \lc
\ee
where $n$ is the density and $\Lambda$ is the cooling rate in erg \cc \s.

The dominant coolants for an evolving cloud change significantly as a function of time.
Early on, the cloud will be largely atomic and ionic in composition and the primary coolant will be CII, while as one moves to regions with at least a few magnitudes of extinction, once molecules have had time to form, CO will dominate the cooling (see Palla and Stahler 2004 for atomic and ionic cooling rates and Goldsmith 2001 for molecular rates).  
The cooling rate per unit volume is highest at early times, and the cooling time is shortest.
The cooling rate drops steadily as the cloud cools and the characteristic cooling time lengthens.
For a representative situation in a cloud well evolved towards molecular steady state (density = 10$^3$ \cc\ and T = 30 K), the molecular cooling rate is
1.5x10$^{-23}$ erg \cc \s (Goldsmith 2001) and $\tau_{cool}$ $\simeq$ 10$^4$ yr.
This is still far shorter than the HI $\rightarrow$ \h2 conversion time.  
Thus, assuming that the cloud immediately attains its steady state density and temperature
configuration should be sufficient for predicting the HI line profiles at all but the earliest times.

An interesting aspect of the cloud thermal evolution is the heating provided by
the initial formation of \h2. 
As discussed by \cite{flower1990}, this results in a secondary temperature plateau
following the initial compression and cooling of the cloud.  
The \h2 formation heating phase ends when the steady state abundance of \h2
is attained; the subsequent heat input from \h2 formation is a part of the cosmic
ray heating rate which provides the steady state gas temperature of $\simeq$ 10 K 
in well--shielded regions.
However, the \h2 heating phase has a duration given by the HI to \h2 conversion time $\tau_{HI \rightarrow H_2} = 2.6\times 10^9/n_0$ yr, where $n_0$ is the proton density in \cc (Paper II).  
The cooling time is thus very short compared to the \h2 heating phase, so once
the latter is completed, the final cool down is very rapid.\
Given the much more rapid conversion of HI to \h2 in the central portion of the cloud
than at the edge, and the accompanying more rapid subsequent cooling, 
the evolving cloud will thus assume a cold--core/warm--edge configuration 
within $<$ 10$^5$ yr, although the temperature distribution will continue to
evolve, especially in the outer portion of the cloud.

In this paper we do not solve for temperature and density profiles. 
Rather, these are chosen to follow those describing a slab illuminated on one side by the standard interstellar radiation field (Le Bourlot et al.\ 1993).  
That cloud model includes photoelectric and cosmic ray heating, and cooling by atomic fine
structure and molecular rotational transitions.  
We have two model clouds which are described in Table \ref{model_common} and
Table \ref{model_specific}.  
The parameters of these clouds are chosen to reproduce those of the clouds studied in Paper II, while the density structure mimics that typically found for small dust clouds (Arquilla \& Goldsmith 1985; Yun \& Clemens 1991; Strafella et al. 2001).
Both models are characterized by a proton column density equal to 10$^{22}$ \c2; the difference is basically the central density, and since the form of the density variation is the same, the sizes  of the clouds differ as indicated in Table \ref{model_specific}.
The proton density and temperature for Model 2 as a function of proton column density measured from the outside of the cloud are shown in Figure \ref{lowdens_cld_vs_coldens}.
A more realistic cloud seen in absorption against the
Galactic background comprises a back--to--back pair of such slabs.  
The simplification in calculating the photodestruction rate by considering only photons incident from a single boundary should produce only a small error in the \h2 photodissociation rate since this is either negligible or dominated by the photon flux from the nearer boundary.
The double--slab model is used to calculate the spectra presented in \S\ref{cloud_spectra}.

\clearpage
\bf
\includegraphics[scale=0.8]{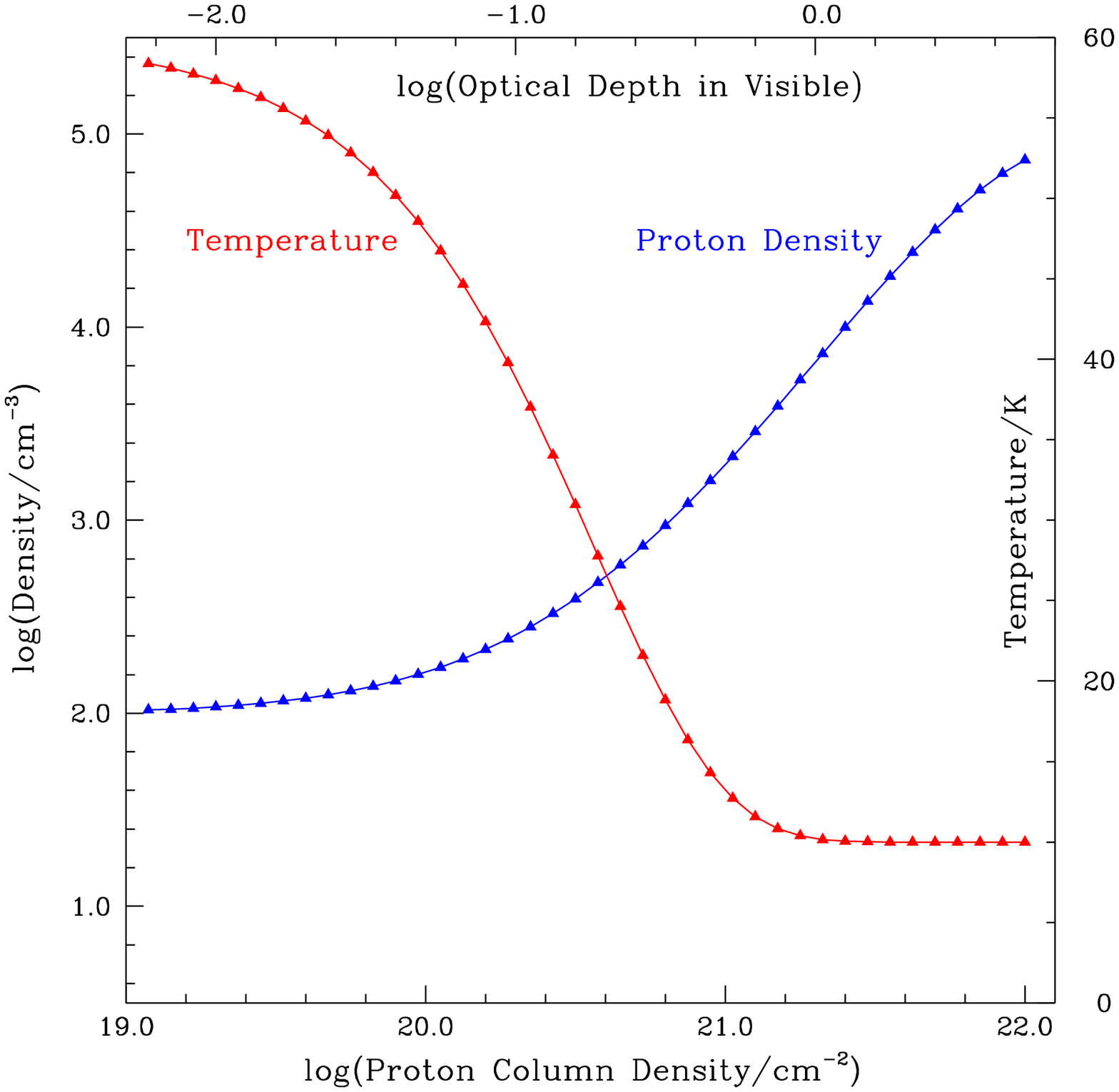}
\caption{\label{lowdens_cld_vs_coldens} Variation of Model 2 cloud parameters as function of displacement from the edge of the cloud, measured in terms of the proton column density.  This model has a higher central density and a smaller physical size than does Model 1.
The full cloud consists of two such slabs back--to--back, with hot edges and a cold center.
The central region is at a temperature of 10 K; the temperature rises when dust attenuation of the external radiation field drops resulting in the photoelectric heating rate exceeding that from cosmic rays (see Le Bourlot et al. 1993 for details).
}
\ef 
\clearpage


\begin{deluxetable}{ll}
\tablewidth{0pt}
\tablecaption{\label{model_common} Parameters of Both Cloud Models}
\tablehead{ \colhead{Parameter} 	& \colhead{Value} 
}
\startdata
Total Proton Column Density\tablenotemark{a}	&	1.0$\times$10$^{22}$ cm$^{-2}$\\
Central Temperature				&	10 K\\
Central HI Thermal Line Width\tablenotemark{b}	&	0.29 km s$^{-1}$\\
Central Total Line Width\tablenotemark{c}	&	0.45 km s$^{-1}$\\
Edge Proton Density				&	103 cm$^{-3}$\\
Edge Temperature				&	58 K\\
\enddata

\tablenotetext{a}{Values for one--sided slab; for a symmetric two--sided slab model the total
proton column density is 2.0$\times$10$^{22}$cm$^{-2}$, corresponding to visual extinction of
10 mag.}
\tablenotetext{b}{one dimensional rms line width}
\tablenotetext{c}{one dimensional rms combined thermal and nonthermal line width}
\end{deluxetable}



\begin{deluxetable}{cccccc}
\tablewidth{0pt}
\tablecaption{\label{model_specific} Model--Specific Parameters}
\tablehead{	 \colhead{Model} 	&\colhead{n$_0$(central)}				&\colhead{Size\tablenotemark{a}} &\colhead{Size\tablenotemark{a}}	&\colhead{Edge Line Width\tablenotemark{b}}
& \colhead{Sphericalized Mass\tablenotemark{c}} \\
		 \colhead{}		&\colhead{cm$^{-3}$}		
		&\colhead{cm}		&\colhead{pc}
		&\colhead{km s$^{-1}$}	&\colhead{\Ms}
}
\startdata
1	&1.6$\times$10$^4$	&4.5$\times$10$^{18}$	&1.46 	&1.03	&108\\
2	&7.4$\times$10$^4$	&2.8$\times$10$^{18}$	&0.91	&0.94	&20\\
\enddata

\tablenotetext{a}{distance from center to edge of one--sided slab cloud}
\tablenotetext{b}{one dimensional rms combined thermal and nonthermal line width}
\tablenotetext{c}{mass of spherical cloud having same density structure as slab used in
the present work}

\end{deluxetable}
\clearpage
  
The density, temperature, and column density profiles as a function of distance from the center of the cloud are shown in Figures \ref{lowdens_cld_vs_radius} and \ref{hidens_cld_vs_radius}.  
Note that since these are one-sided slabs, the equivalent column density for a two-sided cloud would be double this value, corresponding to $N_o$ = 2$\times$10$^{22}$ \c2\ and a total visual extinction of 10 mag \footnote{
The HI column density is converted to reddening using the result of \cite{bohlin1978}, together with the average total to selective extinction ratio determined by \cite{wegner1993}.    
There is relatively small scatter around the mean value quoted in the latter study
(3.1 $\pm$ 0.05) but $R_v$ does show significant variations relative to this mean value when studied in dense regions, so that the conversion of proton column density to visual extinction could well be uncertain by a factor of 2.
}
.
In the following discussion we will use the term ``radius'' to describe the distance from the high--extinction boundary of the one--sided slab, since this would be the ``center'' of the symmetric double slab cloud model.
The density profiles for both models have a constant density core surrounded by an approximately $r^{-2}$ envelope.  
As well as reproducing the observational results described above, this form is similar to the predictions of the isothermal Bonner--Ebert sphere, as given by the approximate solution of Tafalla et al. (2004).  
%
\clearpage
\bf
\includegraphics[scale=0.8]{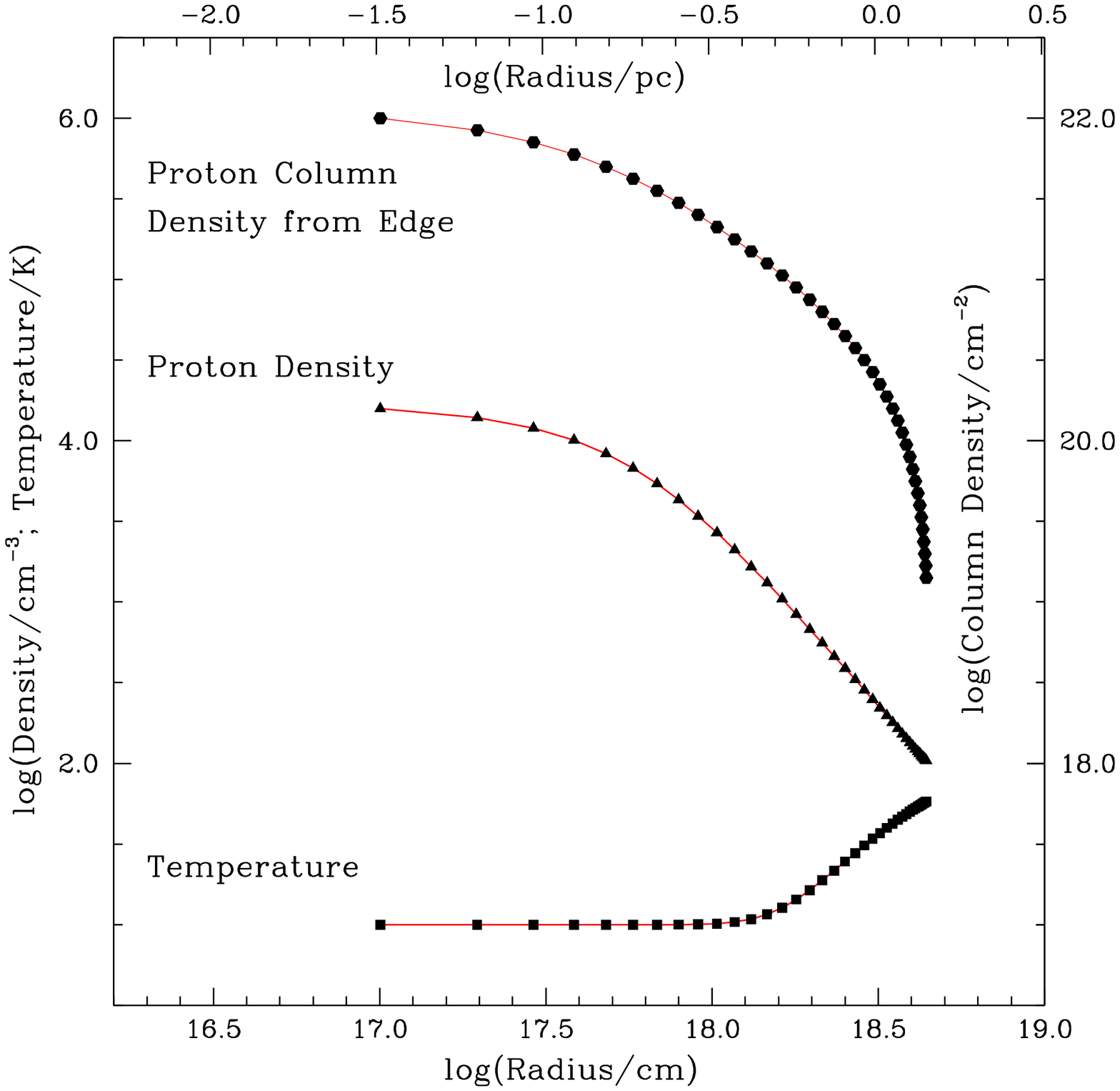}
\caption{\label{lowdens_cld_vs_radius} Variation of Model 1 cloud parameters as function of distance from the center of the cloud.  This model has the lower central density and the larger size.
}
\ef 



\bf
\includegraphics[scale=0.8]{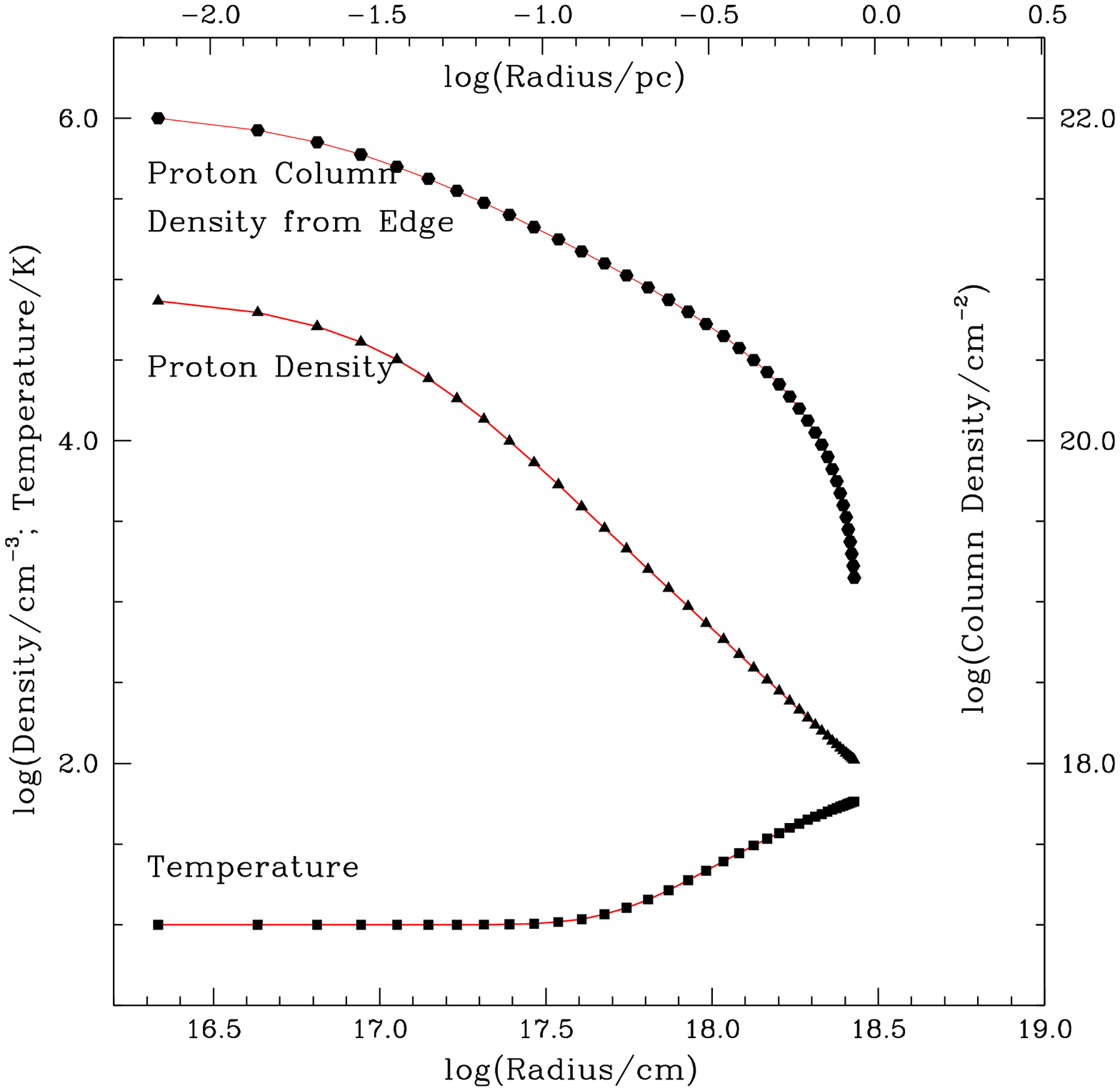}
\caption{\label{hidens_cld_vs_radius} Variation of Model 2 cloud parameters as function of distance from the center of the cloud.  This model has the higher central density and the smaller size.
}
\ef 
\clearpage


The line width is a function of position within the cloud due to variation of the kinetic temperature which is shown in Figures \ref{lowdens_cld_vs_radius} and \ref{hidens_cld_vs_radius}.  
In addition, we assume that there is a nonthermal contribution to the line width varying as 
\be
\delta v_{nonthermal}(r) = \delta v_{nonthermal}(0) (r/r_0)^\beta \lc
\ee
where the line widths are one dimensional rms line widths,
$\delta v_{nonthermal}(0)$ is the nonthermal line width at the center of the cloud,
$r$ is the distance from the center of the cloud, and
$\beta$ is a parameter which we have set to 0.4. 
This relationship has been adopted to reproduce the generally--observed trend of line width increasing with cloud size (e.g. Goodman et al. 1993).
The nonthermal line width is 0.34 \kms\ at the center of the cloud which, as indicated in Table \ref{model_common}, when added in quadrature to the thermal line width at 10 K results in a rms line width of 0.45 \kms.  
These line widths enter very weakly into the determination of the self--shielding rates for \h2\  as well as directly into the radiative transfer in the 21cm line.

\section{TIME--DEPENDENT CALCULATION OF HI and \h2 DENSITIES}
\label{results}
In order to calculate the time--dependent evolution of the atomic and molecular hydrogen densities, we adopt the purely rate--equation approach described in Paper II.  
The intent is to include the key processes that determine the abundance of atomic and molecular hydrogen in a cloud model.  
In addition to cosmic ray destruction, we include photodissociation of \h2, together with self--shielding.   
We use simplified model clouds which have a plausible density variation in only one dimension.  
The physical conditions within the cloud are fixed; only the abundances of atomic and molecular hydrogen vary with time.
This allows us to calculate, in particular, the HI density as a function of time and position within the cloud, which is the basic input for the calculation of spectra line profiles described in Section \ref{cloud_spectra}.

We adopt the rate derived in Paper II for \h2 formation on grains,
$k'_{H_2} = 1.2\times10^{-17}(T/10~$\rm K$)^{0.5}$ cm$^3$s$^{-1}$,
and the \h2 cosmic ray destruction rate $\zeta_{CR} = 5.2\times10^{-17}$ s$^{-1}$.
The position--dependent photodestruction of \h2 is of critical importance in determining the distribution of the species throughout the cloud. 
To treat this we use the shielding function derived by
\cite{draine1996} and given as their equation 37.
We adopt an unshielded photodissociation rate $\zeta_{DISS}(0)$ equal to 1.0$\times$10$^{-11}$ s$^{-1}$.  
This is reduced from what may be considered the ``standard'' value by about a factor of 3 to account for the fact that the type of dense clouds we are modeling are typically embedded in more tenuous extended material which will provide some shielding.  
The results of the calculation do not depend sensitively on this value in any case, as the shielding parameter has to drop to below 10$^{-4}$ to reduce the fractional abundance of atomic hydrogen significantly so that a large change in the unshielded value $\zeta_{DISS}(0)$ produces only a very slight shift in the location of a specific value of $n(HI)/n(H_2)$.

The rate of change of the \h2 density is the difference between the formation and destruction rates.  The former is proportional to the product of the atomic hydrogen density and the grain density (and thus the total proton density).  The molecular  destruction rate is proportional to the \h2\ density, with which we can write 
\be
\label{timedep}
\frac{dn_{H_2}}{dt} = k'_{H_2}n_{HI}n_0 - \zeta_{H_2}n_{H_2} \lc
\ee
with the total proton density defined by
\be
\label{protden}
n_0 = n_{HI} + 2n_{H_2} \lc
\ee
and the total dissociation rate $\zeta_{H_2}$ equal to the sum of the cosmic ray dissociation rate $\zeta_{CR}$ and the photodissociation rate:
\be
\label{destrate}
\zeta_{H_2} = \zeta_{CR} + \zeta_{DISS} \lp
\ee
The photodissociation rate $\zeta_{DISS}$ is given by equation (40) of
\cite{draine1996}, which includes both self--shielding and attenuation of the dissociating radiation field by dust.  In addition to the value of $\zeta_{DISS}(0)$ given above we have adopted a ratio $\tau(1000$\r{A}$)/A_v$ = 5.5,
following discussion in \cite{draine1996}.

The cloud was divided into 40 layers, and equation \ref{timedep} was solved using the routine {\it odeint} from \cite{press1992}, with accuracy parameter set to 2.0$\times$10$^{-6}$.
For each time step the column density of \h2 between each layer and the surface
of the cloud was calculated and used to determine the contribution of the \h2 self--shielding to the shielding function \footnote{
In principle the self--shielding depends on the line width (the $b$ parameter employed by Draine and Bertoldi 1996).  
The combination of thermal and nonthermal line width in the models used here changes by a factor somewhat greater than two from cloud center to edge.
However, the line width has an appreciable effect on the self--shielding only 
for \h2 column densities $\le$ 10$^{18}$ \c2.
Even in the outermost layer, for times after few$\times$10$^{13}$ s, the self--shielding is essentially independent of the line width in the region.
Our use of the line width within a given layer to calculate the self--shielding by
\h2 external to that layer thus results in a negligible error.
}.
There were no problems with the behavior of the solution; a very low level of numerical noise, with fluctuations in the atomic hydrogen density of $\pm$ 5 \% in the steady state solution for the densest portions of the cloud, with $n_{HI}$ approximately equal to 10$^{-4}n_0$, was sometimes visible.

The behavior of the atomic hydrogen density for the two model clouds is shown in
Figures \ref{h1h2_lodens} and \ref{h1h2_hidens}.  
There is reason to be cautious about the detailed behavior at the earliest of 
the times shown here due to the approximation made in this model of an instantaneous evolution to centrally condensed structure with high central density and low temperature.  
Compared to a more realistic model in which the temperature decreases as the density increases, there are two effects:  (1) underestimating the temperature underestimates the \h2 formation rate, which varies as $T^{0.5}$; 
(2) overestimating the density overestimates the \h2 formation rate which
varies as $n_0$ (e.g. Paper II).
These two effects will to a significant extent cancel, but it is still
possible that we are somewhat overestimating the rate of conversion of HI to H$_2$, since it is likely that effect (2) is greater than 
effect (1).

With this modest caveat, we see the same quite striking behavior in the evolution of both high density and low density clouds.
Some of the highlights are:

\begin{itemize}

\item There is a wave of \h2\ formation and accompanying disappearance of atomic hydrogen, which begins in the core of the cloud and propagates outwards

\item The dependence of the HI$\rightarrow$H$_2$ conversion time scale on density results in the central HI density dropping to close to its steady state value while the situation in the outer parts of the cloud has hardly evolved at all.

\item The cloud core reaches its steady state value of $n(HI)$ after $\simeq$ 3$\times$10$^{13}$ s (1$\times$10$^6$ yr) for the low density cloud (Model 1), and after $\simeq$ 1$\times$10$^{13}$ s (3$\times$10$^5$ yr) for the high density cloud (Model 2).

\item When the entire cloud has reached a steady state situation, there is a substantial core with $N_0$ $\geq$ 10$^{21}$ \c2 which has HI density equal to that defined by the equilibrium between \h2 destruction by cosmic rays and formation on grain surfaces.

\end{itemize}

For a region in which $k'_{H_2}n_0 \gg \zeta_{H_2}$, the rate of \h2\ formation exceeds that of \h2\ destruction until a steady state is reached in which $n(HI) << n(H_2)$.  
From equations \ref{timedep} and \ref{protden}, this steady state HI density is given by
\be
n_{HI} = \frac{\zeta_{H_2}}{2k'_{H_2}} \lp
\ee
For the central region of the cloud, the attenuation of the UV radiation field provided by the dust alone is $f_{shield} = e^{-27.5}$ (for $N_0$ = 10$^{22}$ \c2, $A_V$ = 5 mag, and $\tau$(1000 \AA) = 27.5).  
This is sufficient to reduce the photodissociation rate to a level well below that from cosmic rays, even for a purely atomic cloud with no \h2 self--shielding.    
Thus, the central HI density drops to a steady state value
\be
\label{shielded_ss_HI_dens}
n^{*}_{HI} = \frac{\zeta_{CR}}{2k'_{H_2}} \lc
\ee
which with the parameters given above is equal to 2 \cc, even while the outer portions of the cloud have far less \h2 and thus offer relatively little self--shielding for the inner regions until a relatively late time in the evolution of the cloud.

During the moderately early phase of cloud evolution, the HI volume density has a peak at a particular radius.
This is a result of the interplay of the density which is increasing as a function of distance from the cloud edge, and the conversion to molecular form, which occurs more rapidly in the inner portion of the cloud, as discussed above.
At later times, the density of atomic hydrogen drops monotonically as function of distance from the cloud edge, due to the increasing shielding by dust and by \h2 self--shielding.

\clearpage
\bf
\includegraphics[scale=0.8]{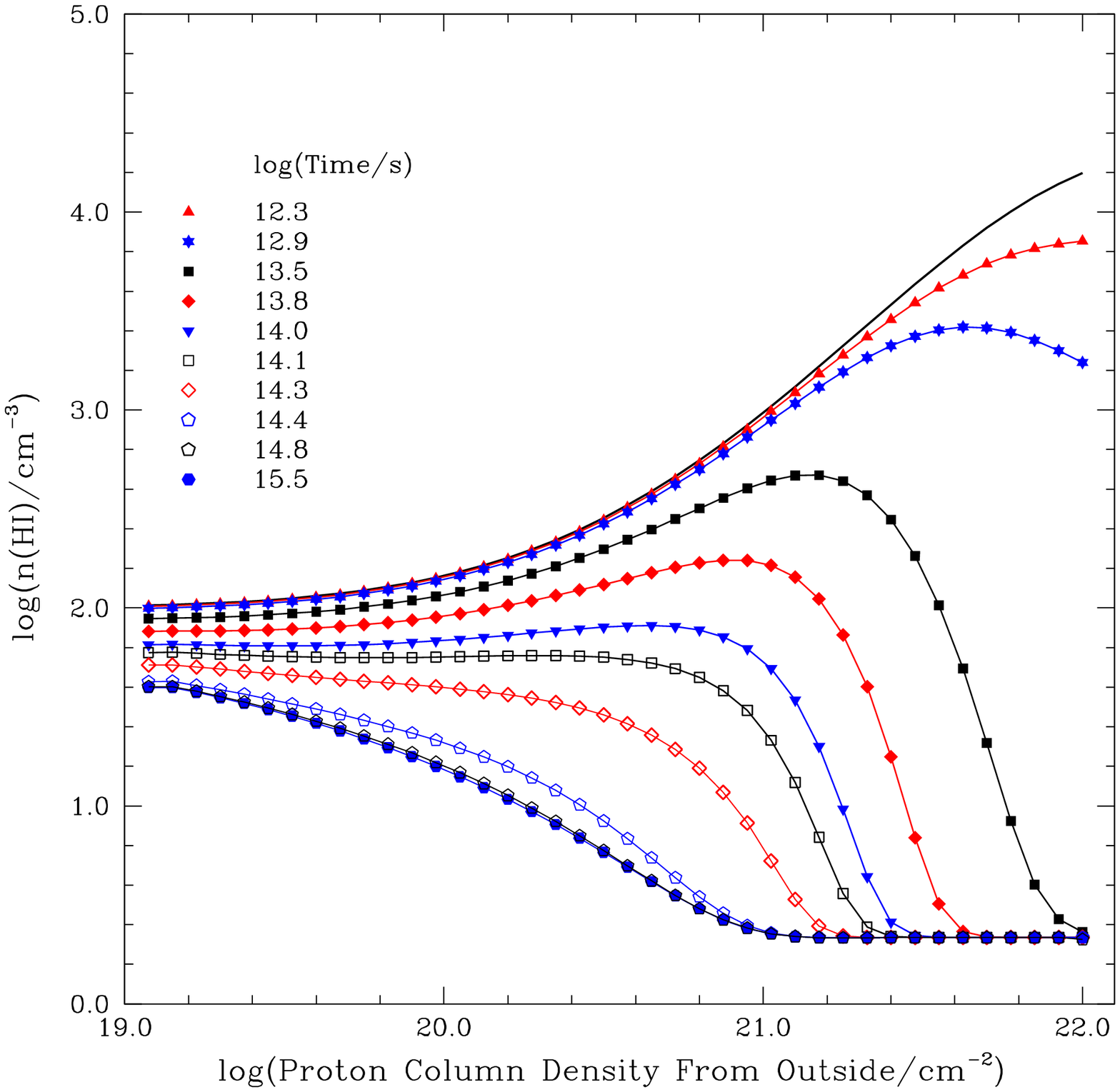}
\caption{\label{h1h2_lodens} HI density as function of proton column density measured from the outside of the cloud at different times between 6.4$\times$10$^4$ yr and 1$\times$10$^8$ yr for Model 1, having $n_0(central)$ = 1.6$\times$10$^{4}$ \cc.  
At the beginning of the calculation ($t$ = 0), the cloud is purely atomic hydrogen throughout, with density indicated by the solid black curve without any symbols.  
The HI disappears first in the center of the cloud due to the high density present there; by $t$ = 3$\times$10$^{13}$ s (1$\times$10$^6$ yr), the HI density in the cloud core has dropped to its steady state value of $\simeq$ 2 \cc.
Starting at this time, a wave of HI disappearance propagates from the center of the cloud to its edge, where the HI $\rightarrow$ \h2 conversion is essentially complete at $t$ = 3$\times$10$^{14}$ s (1$\times$10$^7$ yr).
}
\ef

\bf
\includegraphics[scale=0.8]{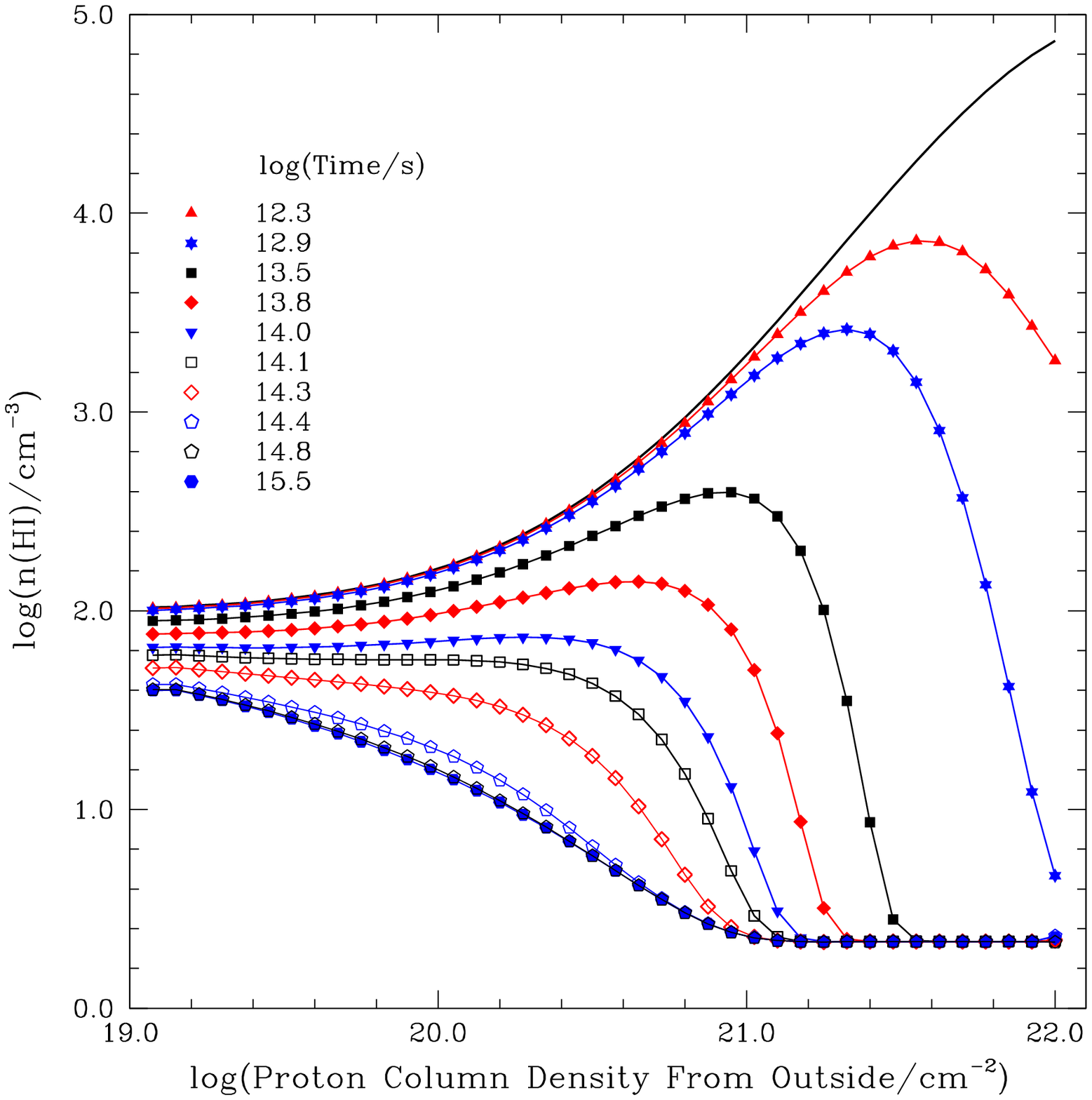}
\caption{\label{h1h2_hidens} As in Figure \ref{h1h2_lodens}, but for Model 2 (high density) cloud having $n_0(central)$ = 7.4$\times$10$^{4}$ \cc, again at times between 6.4$\times$10$^4$ yr and 1$\times$10$^8$ yr.  
The time scale for disappearance of HI in the central portion of the cloud is shorter by a factor equal to the density ratio and thus is $\simeq$ 10$^{13}$ s (3$\times$10$^5$ yr).
The total density in the outer portion of the cloud is the same as in Model 1, and the evolution of the HI density there is characterized by a similar time scale.
}
\ef 
\clearpage

In Figure \ref{frab_lodens} we show the variation of fractional abundance of HI as a function of position within the cloud and of time.
In steady state even the outermost layer of the model clouds has a significant amount of \h2, with  
$n_{H_2}/n_{HI}$ $\simeq$ 0.75. 
This is a result of the very effective self--shielding of the \h2, especially
for the very narrow line widths that we are dealing with here.  
For a column density of \h2 equal to 10$^{19}$ \c2, the self--shielding function
has dropped to a value $\simeq$ 2$\times$10$^{-4}$, while the dust shielding function is essentially unity.
This is sufficient to reduce the \h2 photodestruction rate to $\simeq$ 2$\times$10$^{-15}$ \s, comparable to the formation rate at a proton density equal to 10$^2$ \cc, if $n_{H_2} \simeq$ $n_{HI}$. 

\clearpage
\bf
\includegraphics[scale=0.8]{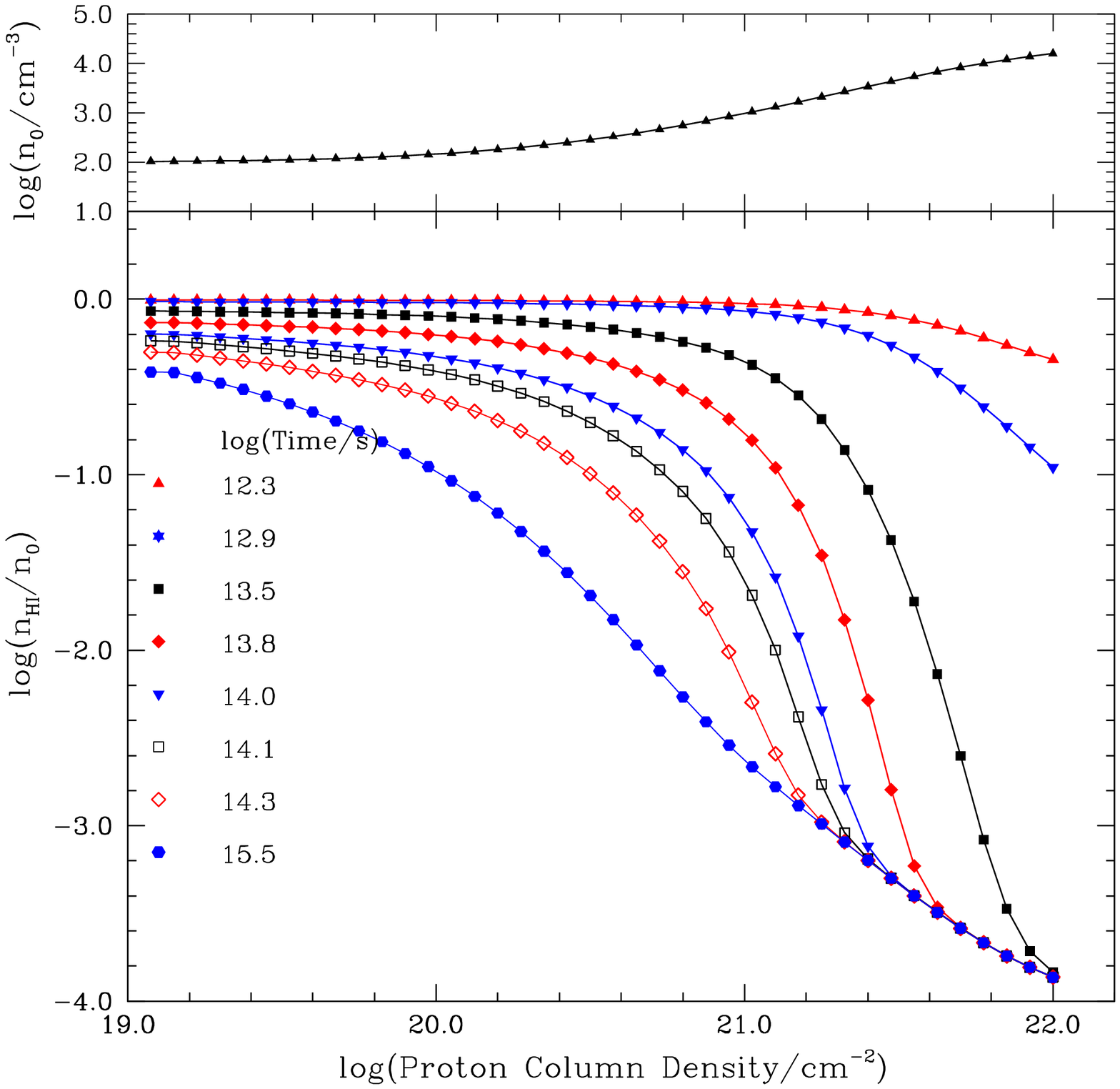}
\caption{\label{frab_lodens} Fractional abundance of atomic hydrogen as a function of proton column density from the surface of Model 1 cloud at times between 2$\times$10$^{12}$ s (6.4$\times$10$^4$ yr) and 3$\times$10$^{15}$ s (1$\times$10$^8$ yr).  
The upper panel shows the proton density as a function of position.  
The lower panel shows the ratio $n_{HI}/n_0$ as a function of position for different selected times during the evolution of the cloud.  
}
\ef 
\clearpage

The drop in the fractional abundance of atomic hydrogen in the outer layer of the cloud as the cloud evolves is only a factor of a few, very modest compared to that in the well--shielded inner portion of the cloud, where the abundance drops by a factor on the order of 10$^4$.
The time scale for HI to \h2\ conversion (see Paper II for additional details) is given by
\be
\label{timescale}
\tau = \frac{1}{2k'_{H_2}n_0 + \zeta_{H_2}} \lp
\ee
As indicated in equation \ref{destrate}, the total \h2 destruction rate is the sum of the cosmic ray and photo destruction rates.
Photodestruction, even with self--shielding, is the dominant destruction pathway for \h2 in the outer portion of the cloud, and being comparable to the formation rate,
contributes modestly to the time scale, which for our model cloud is $\tau_{edge}$ = 1.3$\times$10$^{14}$ s = 4.1$\times$10$^6$ yr.
In the central region of the cloud the time scale is determined entirely by the
first term in the denominator of equation \ref{timescale}, and for model 1 with $n_0$ = 1.6$\times$10$^4$ \cc, we find that $\tau_{center}$ = 2.6$\times$10$^{12}$ s = 0.83$\times$10$^5$ yr.
In the central region of the cloud, a time approximately equal to 10$\tau_{center}$ is required to reach steady state due to the large reduction in $n_{HI}$ that takes place, while at the cloud edge, only about 2$\tau_{edge}$ are required to approach steady state.  
These different time scales and times to reach steady state play critical roles in determining the evolution of the HI 21cm line profiles which we discuss in the following section of the paper.

The evolution of the column densities of atomic and molecular hydrogen for the two cloud models is shown in Figure \ref{coldens_timedep}. 
For Model 1 with $n_0(central)$ = 1.6$\times$10$^4$ \cc\ the HI and \h2 column densities  are equal at time $t$ = 7$\times$10$^{12}$ s (2.2$\times$10$^5$ yr), but this situation is reached in only 1.7$\times$10$^{12}$ s (5.4$\times$10$^4$ yr) for Model 2 with $n_0(central)$ = 7.4$\times$10$^4$ \cc.  
This essentially reflects the ratio of the central densities, which determines the overall time scale in the region of the cloud which is dominant in terms of number of protons.  
The HI column density has fallen to within 10 percent of its final value in time $t$ = 4$\times$10$^{14}$ s (1.3$\times$10$^{7}$ yr) for Model 1 and 3.9$\times$10$^{14}$ s (1.2$\times$10$^{7}$ yr) for Model 2.
These times are not very different since the number densities in the outer regions of the two models are quite similar.  It is in this region that the time scale for HI to \h2 conversion is longest, and hence it dominates the global time scale for this process.

\clearpage
\bf
\includegraphics[scale=0.8]{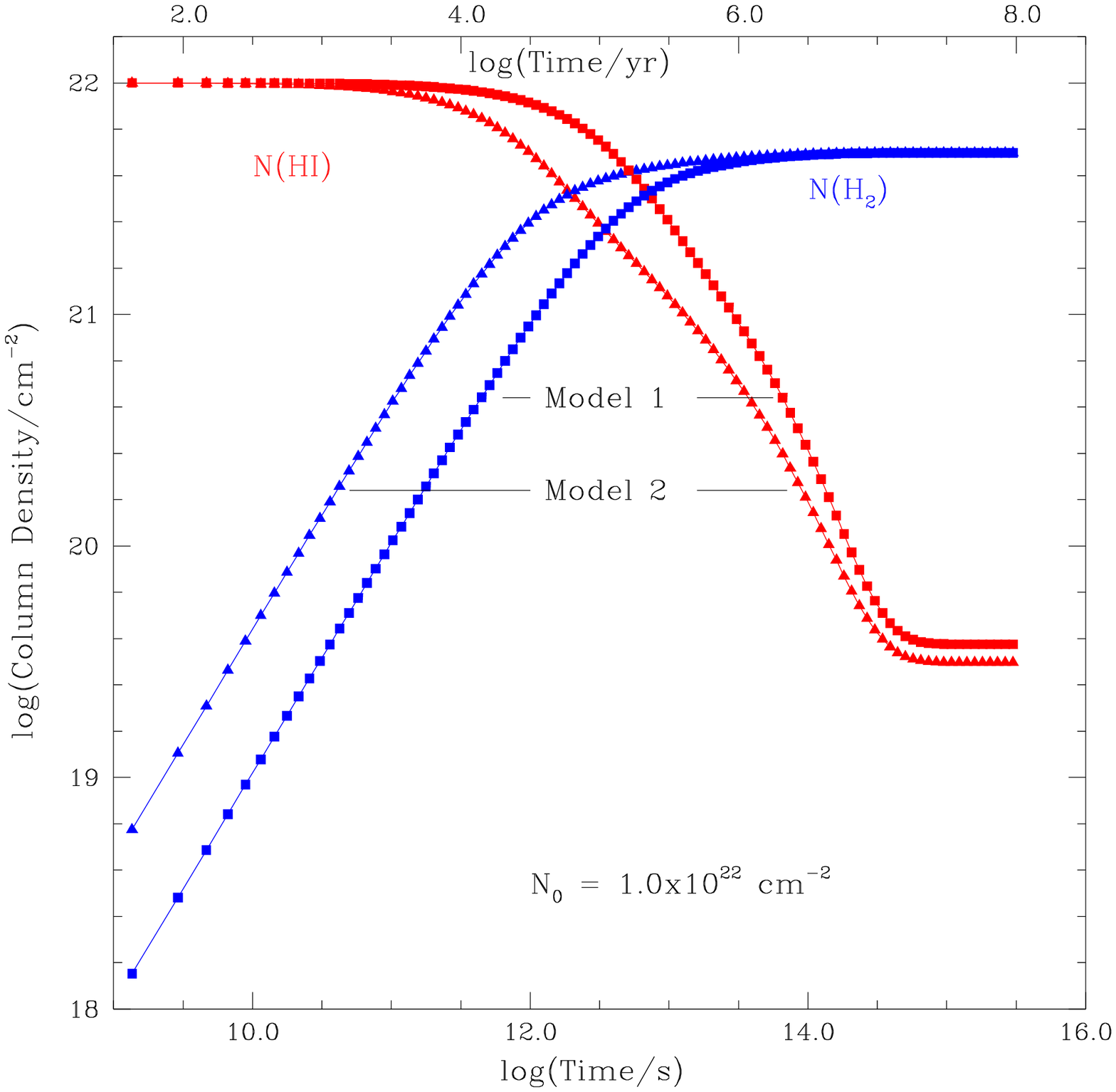}
\caption{\label{coldens_timedep} Variation of the column densities of atomic and molecular hydrogen integrated through single--sided cloud models with central densities given in Table \ref{model_specific}.  
For the lower density Model 1 cloud, the HI and \h2 column densities  are equal at time $t$ = 7$\times$10$^{12}$ s (2.2$\times$10$^5$ yr), but this situation is reached in only 1.7$\times$10$^{12}$ s 5.4$\times$10$^4$ yr) for model 2.  
The HI column density has fallen to within 10 percent of its final value in time $t$ = 4$\times$10$^{14}$ s (1.3$\times$10$^7$ yr) for Model 1 and 3.9$\times$10$^{14}$ s 1.2$\times$10$^7$ yr) for Model 2. 
}
\ef 
\clearpage
 
The two order of magnitude difference in timescales is of considerable importance in appreciating the ability of 21cm HINSA observations to determine or at least to set lower limits on the age of clouds in which the HI abundance is very low.
While the cold HI in the dense, well-—shielded cores dominates the narrow absorption lines characterizing HINSA, its presence can easily be confused or completely hidden by  residual HI in the {\it outer} regions of the cloud.
Thus, as discussed further in \S\ref{cloud_spectra}, the time scale for the entire cloud must be considered when analyzing observations, not just the shorter time scale for the dense cloud core.

Table \ref{HI_param} gives the steady state global properties of HI and other parameters of the model clouds.
The clouds are predominantly molecular, with $N_{HI}/N_{H_2}$ = 0.007 when the entire cloud is included.
If we restrict ourselves to the cold HI, defined here as that having $T$ $\leq$ 12 K, there is some difference
between the models, with $N_{HI}(T \leq 12 K)/N_{H_2}$ between 2 and 2.6 $\times$10$^{-4}$.

\clearpage
\begin{deluxetable}{cccccc}
\tablewidth{0pt}
\tablecaption{\label{HI_param} Steady State Properties of HI and Other Parameters of Model Clouds \tablenotemark{a}}
\tablehead{	\colhead{Model} 				&\colhead{Total $N_{H_2}$} 				&
		\colhead{Total $N_{HI}$}			&\colhead{$N_{HI}$ at $T \leq$ 12 K} 			& 
		\colhead{R($T$ $\leq$ 12 K)\tablenotemark{b}} 	&\colhead{R($A_v$ $\leq$ 1 mag)\tablenotemark{c}}	\\
		\colhead{}					&\colhead{cm$^{-2}$}					&	
		\colhead{cm$^{-2}$}				&\colhead{cm$^{-2}$}					&
		\colhead{cm} 					&\colhead{cm}						\\ }
\startdata
1  &5.0$\times$10$^{21}$  &3.9$\times$10$^{19}$  &3.2$\times$10$^{18}$  &1.5$\times$10$^{18}$  &9.5$\times$10$^{17}$\\
2  &5.0$\times$10$^{21}$  &3.2$\times$10$^{19}$  &1.0$\times$10$^{18}$  &4.8$\times$10$^{17}$  &2.8$\times$10$^{17}$\\
\enddata

\tablenotetext{a}{one--sided slab clouds; all values are twice as large for more realistic double--sided slab models}
\tablenotetext{b}{distance from center of cloud to point where the temperature has risen from 10 K to 12 K}
\tablenotetext{c}{distance from center of cloud to point where the visual extinction to cloud edge is equal to 1 mag}

\end{deluxetable}
\clearpage

Additional insight into the steady state distribution of atomic hydrogen within a molecular cloud can be obtained from Figure \ref{coldens_vs_radius}, for a single--sided Model 1 cloud; the HI column density is 3.9$\times$10$^{19}$ \c2, the \h2 column density is 4.98$\times$10$^{21}$ \c2, while the total proton column density is 1.0$\times$10$^{22}$ \c2.
The integrated HI column density increases linearly with distance from the cloud center in the inner portion of the cloud; this is a result of the constant HI density which results from the steady state high--density limit discussed above.
The HI column density increases more rapidly as one includes the outer portion of the cloud because the HI number density there increases rapidly due to the reduced shielding.
Even so, approximately half the column density of atomic hydrogen is located in relatively cold regions at temperatures less than 30 K.  
This cold HI dominates the narrow absorptions spectra due to its greater optical depth per atom resulting from low temperature and narrow line width, as discussed further in the following section, and consistent with the results of Flynn et al. (2004). 
The results for Model 2 are essentially the same, except that the total column density of atomic hydrogen is a factor $\simeq$ 1.2 lower due to the smaller cloud size.

\bf
\includegraphics[scale=0.8]{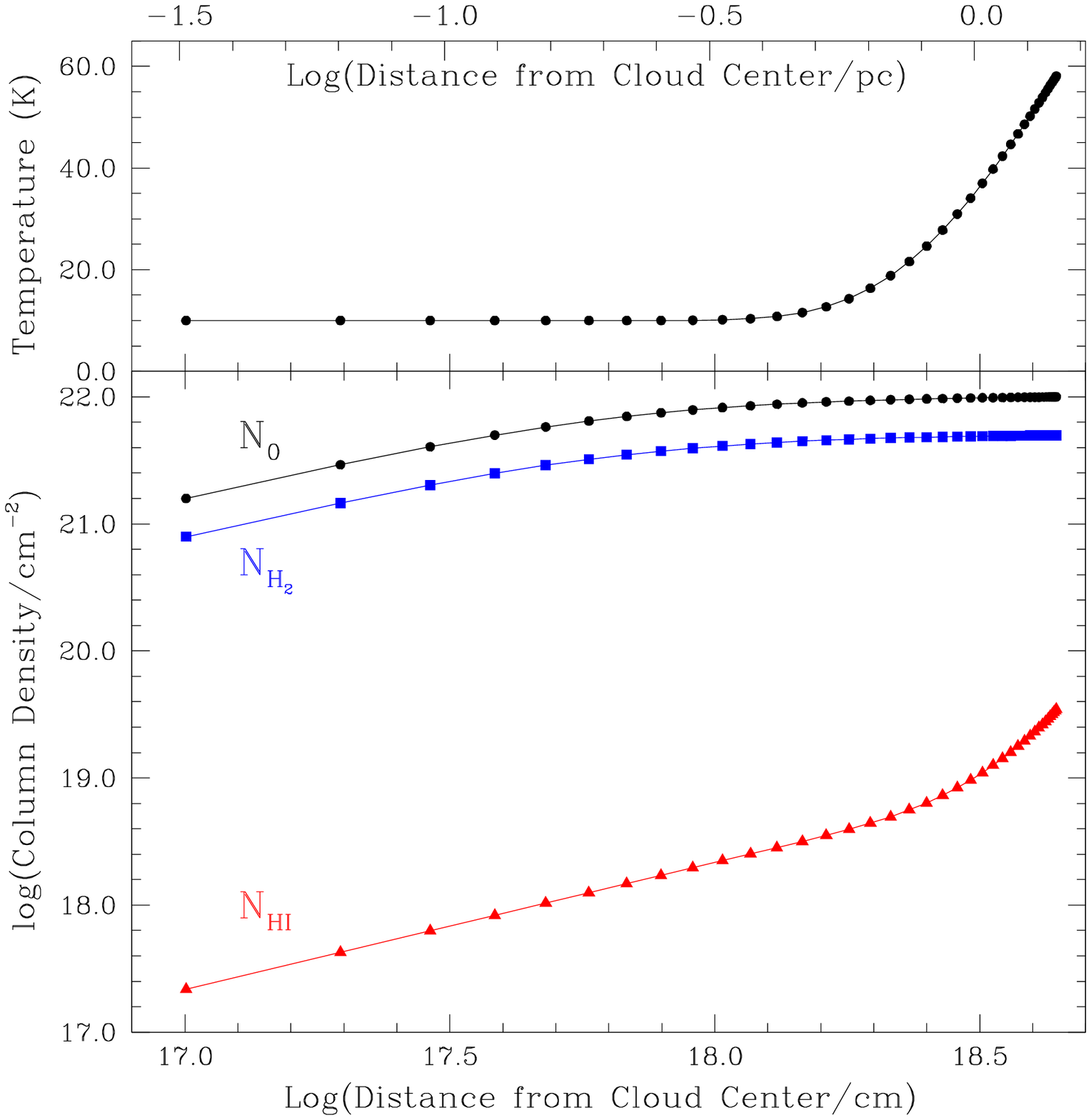}
\caption{\label{coldens_vs_radius}  The lower panel shows the integrated column density of HI, \h2, and protons as a function of distance from the center of Model 1 (low density) cloud under steady state conditions. 
The very low fractional abundance of HI in the high density, well--shielded central portion of the cloud results in the preponderance of the molecular species throughout the cloud.  
As more of the outer portion of the cloud is included, the column density of HI rises more rapidly than that of \h2\ due to the atomic hydrogen density increasing from $\simeq$ 2 \cc\ to $\simeq$ 30 \cc, while the total proton density has fallen by $\simeq$ two orders of magnitude, as shown in Figure \ref{lowdens_cld_vs_radius}.  
The upper panel shows the variation of temperature throughout the cloud. 
}
\ef 


\section{21 CM SPECTRA FROM EVOLVING CLOUDS}
\label{cloud_spectra}

In calculating the HI spectral signature of our model cloud we make a number of simplifying assumptions.  
These include (1) there is a background HI emission source of specified brightness temperature and line width which is uniform over the antenna beam; (2) there is no HI between that source and the model cloud or between the model cloud and the observer), and (3) the model cloud is uniform over the two dimensions defining the antenna beam and varies only in the third, line of sight, dimension. 
In the Rayleigh-Jeans limit and assuming that the level populations are in LTE at the kinetic temperature ($T_{spin} = T_{kinetic}$), we calculate the optical depth within each slab comprising the cloud, and solve the equation of radiative transfer in a stepwise fashion starting from the background.
In order to be somewhat closer to the observations, all of the model spectra presented here are for double--sided clouds, which are two single--sided clouds discussed above combined into a slab cloud with two warm edges and a dense, cold central region.  
There are thus 80 layers in the double--sided cloud, with density, temperature, HI density, and line width defined in \S\ref{model}.
The HI density in each layer for each of the time steps is stored and is used with the background line profile to calculate the optical depth and the emergent line profile as a function of time.

The background line intensity plays a critical role in many situations.  
This is a direct consequence of the fact that the outer layers of the cloud we are studying have temperature comparable to that of the background 21cm emission.
It is thus possible for the foreground cloud to add to the emission while the cold gas in its center simultaneously produces a narrow absorption feature.
There are thus a large number of combinations of cloud models, times, and background emission.  
We here restrict ourselves to presenting representative results that (1) illustrate the range of spectra that can be produced and (2) focus on the time evolution of the HI narrow self--absorption features in order to constrain the time required to obtain results that are broadly consistent with observations.

\subsection{General Results}
\label{general_results}
In Figure \ref{Model1_double_60Kbg_6times} we show the results for a double--sided Model 1 cloud at six different times.
In this case, the background source is modeled as a line having peak temperature of 60 K and a full width to half maximum (FWHM) line width of 30 \kms. 
The central velocity of the background emission is offset by 5 \kms~ from that
of the foreground absorbing cloud.
%
%
\clearpage
\bf
\includegraphics[scale=0.8]{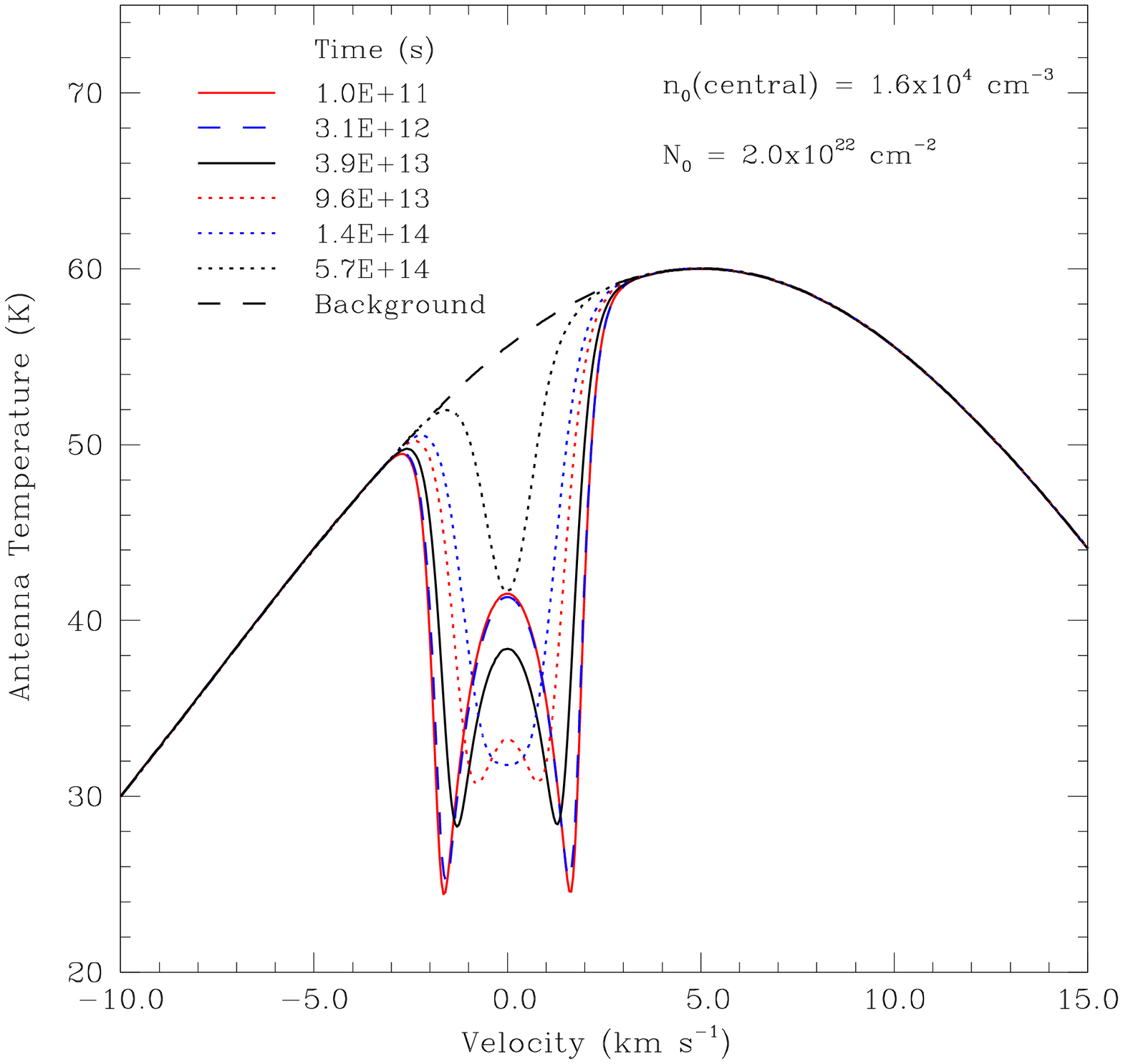}
\caption{\label{Model1_double_60Kbg_6times}  Spectrum of double--sided Model 1 cloud at six different times between 1$\times$10$^{11}$ s (3.2$\times$10$^3$ yr) and 5.7$\times$10$^{14}$ s (1.8$\times$10$^7$ yr). 
The background spectrum produces a peak antenna temperature of 60 K at velocity of 5 \kms, as described in the text.  
Only the spectra at the two latest times ($t$ $\ge$ 10$^{14}$ s or 3$\times$10$^6$ yr) are at all similar to the HINSA profiles shown in Figure \ref{spectra}. 
The spectrum at $t$ = 1.4$\times$10$^{14}$ s (4.5$\times$10$^6$ yr) is somewhat flat--bottomed due to saturation, and is not typical of those observed. This indicates that lower 21cm optical depths, corresponding to times $t$ $\ge$ 3$\times$10$^{14}$ s (10$^7$ yr) are required to be consistent with the observations.
}
\ef 
\clearpage
Several interesting and significant results are evident from this set of spectra.
First, at relatively early times, the very large column density of HI in the foreground produces strong absorption.  
This gas is almost all colder than the  background emission over the velocity range of the foreground cloud (between 50 \kms and 58 \kms).  

The absorption profile is complicated by the temperature gradient within the foreground cloud.  
For example, at time equal to 3.9$\times$10$^{13}$ s (1.2$\times$10$^6$ yr), the total HI column density through the cloud is $\simeq$ 10$^{21}$ \c2 and the peak optical depth $\simeq$ 30.  
At line center one sees in to a layer close to the surface, where the temperature is relatively warm, with $T$ $\simeq$ 40 K.  
As one moves off line center, the optical depth drops, and one sees the temperature deeper into the cloud, where the temperature is lower.  
The optical depth is equal to unity at velocity offsets of $\pm$1.6 \kms, corresponding to a depth in the cloud at which $T$ = 25 K.  
This defines the two sharp minima in the absorption spectrum.
For larger velocity offsets, the absorbing line becomes increasingly optically thin, and the line profile joins that of the background emission.

The column density of HI in the cloud is sufficient for the line to remain optically thick until approximately 10$^{14}$ s (3$\times$10$^6$ yr) have elapsed.
After about 2$\times$10$^{14}$ s (6.4$\times$10$^6$ yr), the column density of HI has dropped sufficiently that we see an optically thin profile.  
The final profile presented here, at time 5.7$\times$10$^{14}$ s (1.8$\times$10$^7$ yr), is essentially that characteristic of the absorbing cloud in a steady state.
In this situation, the peak optical depth of the Model 1 cloud is 0.71.

The line profiles for the earliest times may, as suggested above, be inaccurate in detail since we have assumed that the cloud has instantaneously evolved to its final density and temperature profile.  
Thus, we may not expect to see quite such striking absorption features and peculiar line profiles, but again as discussed previously, the general evolution of the cloud from large HI column density to having a relatively low fractional abundance of atomic hydrogen is certainly robust.  
The different time scales for the HI $\rightarrow$ \h2 conversion in regions of different densities will also generally characterize gravitationally bound clouds which are, in general, centrally condensed. 
What is clearly evident from the spectra shown here is that in order to see profiles similar to those observed (e.g.\ Figure \ref{spectra}), the abundance of HI in the warm envelope must be dramatically reduced relative to that in the initially purely atomic cloud.  
This means that the much longer time scale for the lower density region enters into determining when we can see the very narrow self--absorption profiles which are primarily produced in the colder central regions of the cloud.
A time exceeding 1$\times$10$^{14}$ s (3$\times$10$^6$ yr) is required for the HI $\rightarrow$ \h2 conversion to have proceeded sufficiently far to avoid obvious saturation in the single--dip absorption spectra.  
The exact depth of the absorption profile depends on the details of the cloud structure.
Our conclusions here are consistent with observational results analyzed in terms of the fractional abundance of cold HI in a cloud with a single uniform (average) density (Paper II).  
The present modeling results demand times $\simeq$ 3$\times$10$^{14}$ s, or 1$\times$10$^7$ yr, but since we have not fitted specific HI spectra and determined HI and \h2\ column densities it is not appropriate to give more precise values at this point.

Another manifestation of the drop in the atomic hydrogen density as
the cloud evolves is the reduction of the apparent line width of the absorption profile, which continues through the later part of the time period considered here. 
Considering only the largely Gaussian pure absorption spectra, the line
width decreases from 2.7 \kms\ at time 1.4$\times$10$^{14}$ s (4.5$\times$10$^6$ yr) to 1.4 \kms\ at time 5.7$\times$10$^{14}$ s (1.8$\times$10$^7$ yr).  
The line broadening at the earlier time is essentially a result of saturation.
The width of the optical depth profile, defined as the width to half maximum, increases until time 1.4$\times$10$^{14}$ s (4.5$\times$10$^6$ yr), and then decreases slightly, while the line width at 10\% of maximum increases monotonically throughout the calculation.  
This is basically a result of the changing relative amount of HI in regions having lower and greater velocity dispersion.  
At early times, the large column density of HI in the dense, relatively quiescent core completely dominates the optical depth profile, but as time goes passes, the atomic hydrogen in the exterior layers of the cloud becomes relatively more important.
The result is a FWHM line width of 1.6 \kms, compared to that of 1.05 \kms~ (combined thermal and nonthermal) expected from the core of the cloud alone.
Under close examination, the line profile is not accurately described by a Gaussian profile, since the although the velocity distribution within each layer of the cloud is a Gaussian, the variation of the line width as a function of position results in the line profile no longer being purely Gaussian.  

The corresponding spectra for Model 2 (having higher central density and smaller size) are shown in Figure \ref{Model2_double_60Kbg_6times}.
While the general behavior is similar to that for Model 1, there are several noticeable differences.
The early--time evolution of the line profile is similar to that for Model 1, with very prominent, broad absorption evident, and structure from the temperature gradient clearly visible in terms of the central filling--in of the absorption profile.
However, starting at about 5$\times$10$^{13}$ s (1.6$\times$10$^6$ yr), it is evident that the drop in the column density in the higher density cloud is more dramatic that in the lower density cloud.
By $\simeq$ 1$\times$10$^{14}$ s (3$\times$10$^6$ yr), the Model 2 profile has a single minimum, while that of Model 1 still has a peak within the absorption feature.
This is due to the smaller 21cm optical depth (3.0 for Model 2 compared to 6.8 for Model 1) in the presence of the temperature gradient.
The line profile for Model 2 is broadened by saturation, but narrows and weakens as the HI to \h2 transformation is completed.
In steady state, both clouds have optically thin HINSA features, with $\tau_{max}$ = 0.49 for Model 2
compared to 0.71 for Model 1.
The weaker feature for the denser cloud is really a reflection of its smaller size and hence the shorter path with the steady state HI density given by equation \ref{shielded_ss_HI_dens}.
The optical depth in steady state for Model 2 is closer to those determined from observations of HINSA clouds, typically a few tenths, although a few sources do have $\tau$ approaching unity.  

\clearpage
\bf
\includegraphics[scale=0.8]{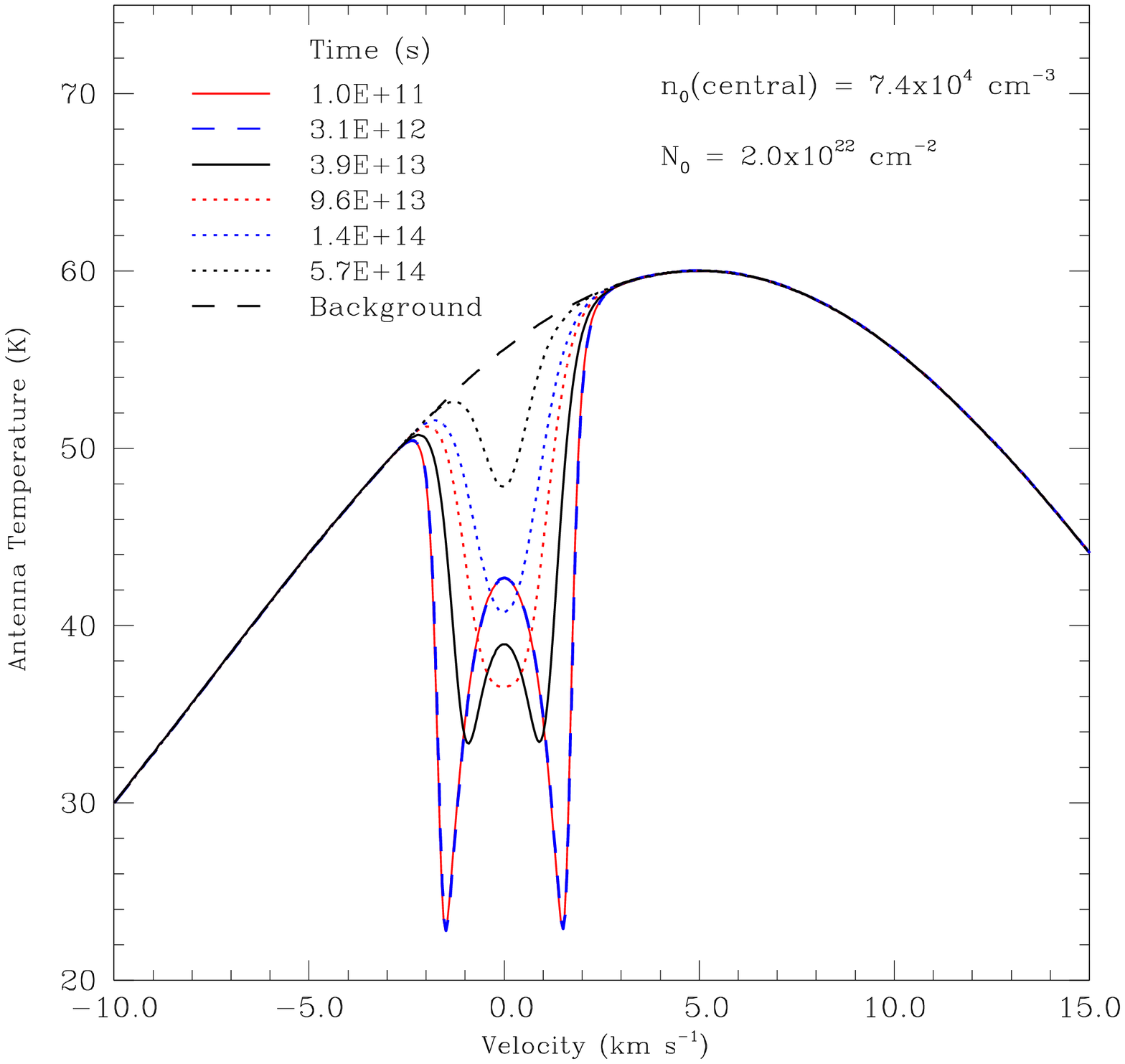}
\caption{\label{Model2_double_60Kbg_6times}  Spectrum of double--sided Model 2 foreground cloud at six different times between 1$\times$10$^{11}$ s (3.2$\times$10$^3$ yr) and 5.7$\times$10$^{14}$ s (1.8$\times$10$^7$ yr). 
}
\ef 
\clearpage


\subsection{Effect of Background Temperature}

As a consequence of the complex temperature structure of the clouds modeled here, the antenna temperature of the background radiation can have a dramatic effect on the line profiles produced by the foreground cloud.  
The temperature of the atomic hydrogen in the well--shielded, cold portion of the cloud is ($\leq$ 20 K) is colder than the antenna temperature of any commonly encountered background signal.
This is not the case for the atomic hydrogen in the warmer, outer portion of the cloud, which can produce line profiles in which the contribution of the cold ``HINSA'' feature is much less obvious than in the preceding discussion.
In Figure \ref{quadplot_3Tbg} we show examples with the peak background temperature reduced from the 60 K used in Figures \ref{Model1_double_60Kbg_6times} and \ref{Model2_double_60Kbg_6times}, to 45 K and 35 K.  
The actual antenna temperature across the range of velocities of the absorbing cloud is lower than this, and varies due to the 5 \kms\ velocity offset combined with the 30 \kms\ background line width.
In each case the spectrum of the background source has been subtracted from the spectrum produced by the model cloud located in front of the background source.  
This allows details of the spectra, which would be lost were the entire background emission spectrum to be plotted, to be seen.  
The spectra in Figure \ref{spectra}, as well as those in Papers I and II, were all taken in total power (on source only) mode, and there is no danger of features being produced by small differences in the background spectrum on-- and off-- the foreground cloud.
Recalling from Figures \ref{lowdens_cld_vs_coldens}, \ref{lowdens_cld_vs_radius}, and \ref{hidens_cld_vs_radius}, the temperature is $\simeq$ 60 K in the outermost layer of the cloud, falls to $\simeq$ 20 K at the point where there is 1 mag of external dust extinction, and drops to 10 K in the central portion of the cloud.  
Thus, for $T_{A~background}$ = 52 to 58 K as produced by a cloud with peak intensity $T_{A~background}$ at 5 kms$^{-1}$ = 60 K, there is very little gas at temperatures comparable to that of the background, so one gets only absorption lines irrespective of the evolutionary state of the (foreground) cloud and the distribution of atomic hydrogen within it.

\clearpage
\thispagestyle{empty}
\bf
\vspace*{-15mm}
\includegraphics[scale=0.8]{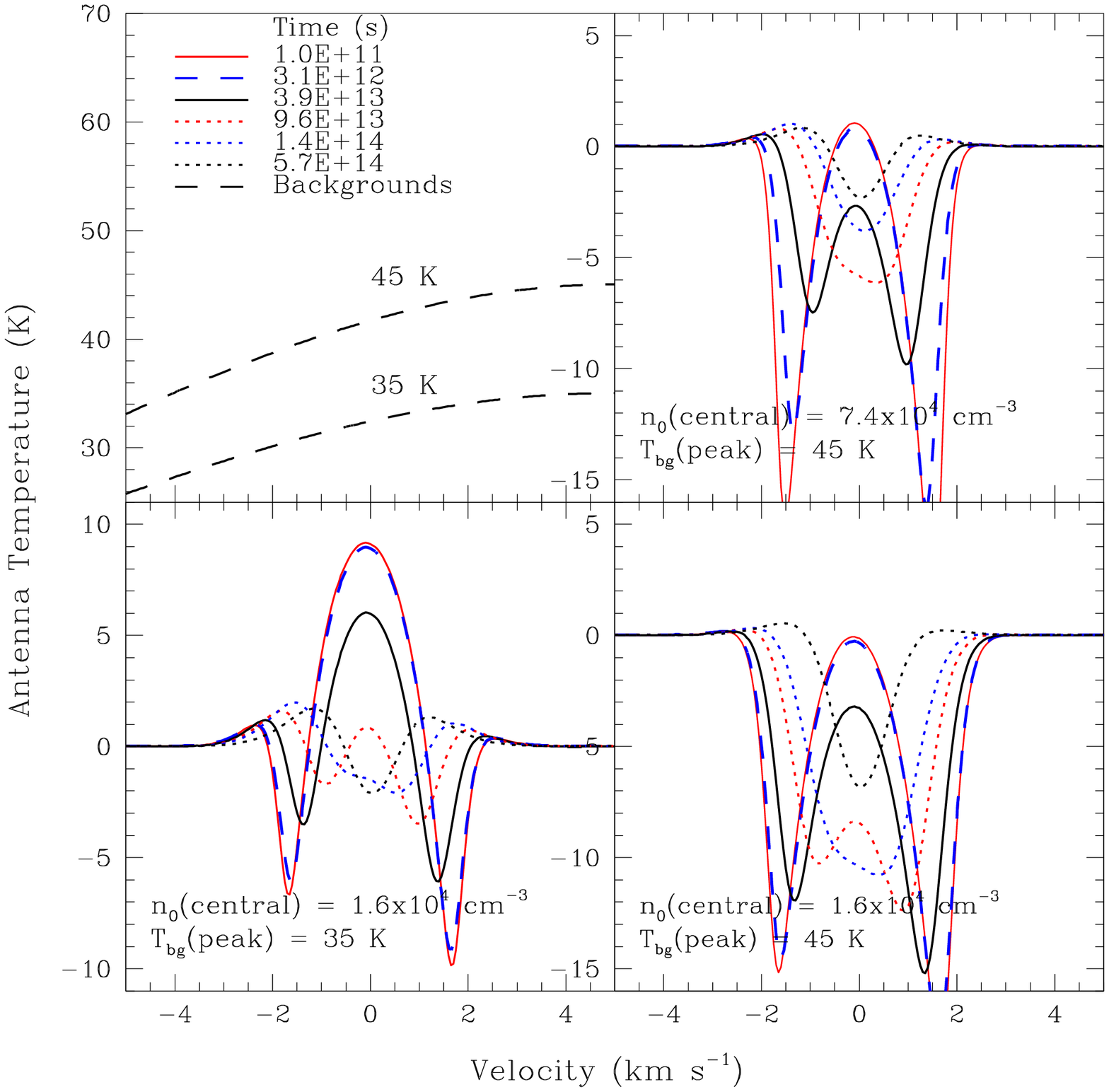}
\caption{\label{quadplot_3Tbg} Representative examples showing effect of different background temperatures on the spectra produced by a foreground cloud at six different times between 1$\times$10$^{11}$ s (3.2$\times$10$^3$ yr) and 5.7$\times$10$^{14}$ s (1.8$\times$10$^7$ yr).
The background spectra have the same FWHM line width (30 \kms) and central velocity (5 \kms) as those in previous two Figures, but the peak temperatures are 35 K and 45 K rather than 60 K.  
For each case, the background spectrum has been subtracted from the observed spectrum that would be produced by the model cloud in front of the background source to improve the visibility of fine details produced by the model cloud.
The spectra of the background source are shown in the upper left hand panel.  
The  key for the different times at which the cloud evolution is sampled applies to the three other panels which
show profiles which are the difference between the observed spectrum
and the background spectrum.
In the upper right is the high density (Model 2) cloud with a background temperature of 45 K,
at the lower left is the low density (Model 1) cloud with a 35 K background, and the lower right
panel shows the low density (Model 1) cloud with a 45 K background.
}
\ef
\clearpage

Once the background temperature falls to $\simeq$ 40 K (for $T_{A~background}$ at 5 kms$^{-1}$ = 45 K), the situation is somewhat different, as seen in the right hand panels of Figure \ref{Model2_double_60Kbg_6times}.  
At early evolutionary times, the 21cm line observed from the exterior of the cloud becomes optically thick at a depth where the temperature is greater than 40 K, and so the antenna temperature at the velocity of the center of the self--reversed absorption profile is slightly greater than that of the background.  
As one moves to the line wings, one sees deeper into the cloud, and $\tau(v)$ = 1 occurs in colder gas so one again sees absorption
\footnote{The line profiles in Figures \ref{Model1_double_60Kbg_6times} and 
\ref{Model2_double_60Kbg_6times} have absorption minima at early times which reach the same value of antenna temperature.  In Figure \ref{quadplot_3Tbg}, which plots the line  minus the background, the absorption minima appear to have different values due to the variation in the antenna temperature of the background across the velocity range of the foreground cloud.
}.
As the cloud evolves, the hydrogen in the outer layers becomes optically thin, and the line profile is dominated by the absorption in the colder central region of the cloud. 
This remains somewhat optically thick even until times $\simeq$ 10$^{14}$ s (3$\times$10$^6$ yr), as revealed by the asymmetry of the line minus background absorption spectra.  
At relatively late times, the 21cm line is optically thin, but there is more atomic hydrogen in the cold core of the cloud than in its envelope, so the net effect at line center is absorption.  
As one looks at the wings of the line profile of the foreground cloud, since the line width in the outer portion of the cloud is broader, the emission is dominated by warm hydrogen in the outer part of the cloud, and is manifest as the small maxima offset by $\pm$ 1 \kms\ from the foreground cloud centroid velocity.
With a peak background antenna temperature of 45 K, this effect is quite modest, and there would not be much difficulty in identifying the absorption feature as due to cold gas, and obtaining a reasonably accurate determination of the column density of HI in the core of the cloud.

The lower left--hand panel of Figure \ref{quadplot_3Tbg} shows the situation for a Model 1 (low density) cloud in the foreground of emission having a 35 K peak antenna temperature.
Here, the situation is much more extreme due to the larger fraction of material at a temperature greater than that of the background.
The profile at early times is dominated by a central emission feature which reaches 10 K above the background antenna temperature at the cloud centroid velocity.
As the HI density drops, the signature of the atomic gas in the foreground cloud becomes much weaker, and during the phase in which the 21cm line is somewhat optically thick ($t$ $\simeq$ 10$^{14}$ s or 3$\times$10$^6$ yr), the line profile is very complex.  This results from the inverted center of the absorption line profile reaching just about to the level of the background, combined with weak net absorption features at $\pm$ 1 \kms\ from line center together with weak excess emission features at
$\simeq$ $\pm$ 1.5 \kms\ from line center.  
This very complex undulatory line profile might easily escape notice as being due to absorption by cold HI -- the wiggles are similar to those seen in the spectrum of B227 in Figure \ref{spectra} at velocities between 4 \kms\ and 14 \kms, although the background temperature of the B227 profile is much larger, with $T_{A~background}$ $\simeq$ 80 K, than we are considering here.
At very late times, we are left with a central absorption feature flanked by modest excess emission, which again would be recognizable as a signature of cold HI.

The overall conclusion here is that while a narrow HI absorption profile at the centroid velocity of a dense cloud defined by molecular tracers is always a signature of associated cold HI, the absence of readily detectable HI absorption does not indicate that there is no cold atomic hydrogen present. 
The availability of a molecular tracer to determine the velocity and line width of the relatively quiescent gas in the central portion of the cloud is highly advantageous for interpreting the HI absorption spectra.
A low background temperature (or equivalently an enhanced temperature of the outer portion of the foreground cloud as would be produced by being in a region of enhanced interstellar radiation field intensity) can produce a line profile which disguises the contribution of the cold absorbing gas.

\section{DISCUSSION AND CONCLUSIONS}
\label{conclusions}

Understanding the processing of material in the interstellar medium, its relationship to star formation, and the timescales for different parts of the cycling between stars and gas is a complex undertaking.  
In the present work we have investigated the transformation of atomic to molecular hydrogen in a region containing gas and dust which has been compressed.
The increased density raises the rate of molecular formation and the larger column density reduces the rate of molecular photodestruction, which together result in a steady state overwhelmingly molecular in composition.  
The present calculation represents a step forward relative to single--zone time dependent models and multizone time--independent models.
By adopting a fixed density and temperature distribution, the present calculation is still an idealization of the evolution of an actual entirely atomic cloud evolving to a molecular central region and atomic/molecular envelope, with the dense core possibly moving to a star forming phase.  

Recent studies of atomic hydrogen in well--defined molecular clouds (Papers I and II) emphasized the widespread occurrence of HI Narrow Self--Absorption, or HINSA in the nearby Taurus molecular cloud complex.  
The ubiquity of cold HI found in earlier studies (e.g. Knapp 1974; see Papers I and II for additional references) in predominantly molecular regions has been confirmed by a more recent survey of a number of regions at varying distances, but still relatively close to the Earth (Kr\v{c}o, Goldsmith, \& Li 2006).  
While these studies are still limited by angular resolution and sensitivity, and the possible effect of atomic hydrogen between the absorbing cloud and the Earth (due to greater distance of some of these sources compared to Taurus),  there is no reason to think that atomic hydrogen is not present in well--shielded molecular cloud cores, with a fractional abundance $n_{HI}/n_{H_2}$ between a few times 10$^{-4}$ and a few times 10$^{-3}$ (Papers I and II)
\footnote{
One apparently discrepant cloud is the globule studied by \cite{klaassen2005} and found to have 0.02 $\leq$ N(HI)/N(H$_2$) $\leq$ 0.25.  However, the conclusion that this 2 kpc--distant cloud is gravitationally unbound suggests that it may be no where near steady state and consequently has a much higher fractional abundance of atomic hydrogen than the clouds studied in detail in Paper II, which are close to virial equilibrium.
}.  
The time required to reduce the density of HI from 100\% of the total proton density in the gas phase, to less than 0.01\% of this value was determined in Paper II to be on the order of a few million years.
This was based on assuming that the compression instantaneously reduced the photodissociation rate to zero and increased the average densities of those clouds to their current average values of a few thousand \cc.    
Since the evolution of the HI and \h2\ densities depends on both of these parameters, the present study was undertaken to include realistic treatment of photodissociation as a function of position in a cloud with structure including density and temperature gradients.  
We do not treat the time dependence of $n$ and $T$, but rather fix them to conform with models representing centrally condensed hot edged predominantly molecular clouds which are largely consistent with observations.  

The present work also attempts to make a closer connection with HINSA observations by calculating the line profiles that would be measured if the model cloud were observed in absorption against the bright, relatively uniform background provided by general Galactic HI emission.  
We have considered two model clouds, treated as slabs which vary in one dimension and which are infinite in the two other dimensions.  
Each cloud has a total proton column density of 10$^{22}$ \c2.
The interstellar radiation field is assumed to be incident on one side of the cloud, which is heated to 60 K.
As one moves away from this interface, the density increases and the temperature drops to 10 K.
The difference between the two models is the central density (1.6 and 7.4 $\times$ 10$^4$ \cc\ for Model 1 and Model 2, respectively),
and the cloud size, which differs so as to result in equal column densities.
Each complete model cloud used to calculate the HINSA spectra comprises two such slabs back to back, and is thus an approximation of
a warm--edged cloud with HI fractional abundance $n_{HI}/n_{H_2}$ $\simeq$ 1 in the outer layers and $\simeq$ 10$^{-4}$ in the center of the clouds in steady state.

The time dependence of the spectral signature of the HI seen in absorption provides an important diagnostic of the evolution of the HI distribution in the cloud.  
The time scale for the conversion of HI to \h2\ in well--shielded cloud cores is inversely proportional to the proton density.
Thus, in the denser central portion of the cloud, the time scale $\tau_{center}$ for HI $\rightarrow$ \h2\ conversion is approximately 10$^{12}$ s (3.2$\times$10$^4$ yr).  
However, during the course of the evolution of the cloud, the HI density drops 3 to 4 orders of magnitude, thus requiring time $\simeq$ 10$\tau_{center}$ = 10$^{13}$ s (3.2$\times$10$^5$ yr).  
If we consider that the central portion of the cloud is heated by the energy released during the HI $\rightarrow$ \h2\ conversion process, then since the characteristic cooling time is much less than that of this heating phase, we end up with a $\simeq$ 10 K cloud core with a very low steady state value $n(H)$ $\simeq$ 2 \cc\ after a time equal to a few $\times$ 10$^{13}$ s ($\simeq$ 10$^6$ yr) following the initial cloud compression has passed.
The time scale in the outer layers of the cloud is far longer due to the lower density there, with $\tau_{edge}$ $\simeq$ 10$^{14}$ s (3$\times$10$^6$ yr). 
 
The 21cm spectra sample atomic hydrogen throughout the evolving cloud.
Since the optical depth in a given region of the cloud varies inversely as its temperature, the optical depth per hydrogen atom is considerably less in the outer portion than in the central part of the cloud.  
The peak optical depth varies inversely as the line width, which being larger in the outer portion of the cloud than in the center, similarly results in a lower absorption per hydrogen atom in the periphery of the cloud than in the center.  
The combination of these two factors explains why the steady state HI column density in the outer part of the cloud is almost invisible compared to that in the center.

The HI density in the low and moderate density regions of the cloud takes the longest to approach its steady state value. 
The evolution of these regions of the cloud thus provides a strong observational constraint on the ``age'' of the cloud. 
At early times, the HI in the low and moderate density regions of the cloud is so abundant that it overwhelms the atomic hydrogen in the cloud core. 
This situation is readily identifiable from the spectra produced:  the large quantity of HI produces (depending on details of the temperature structure) emission features, or absorption features significantly broader and deeper than those observed.
The fact that such spectra are not observed may in part be due to simplifying assumptions made about the time independence of the density and temperatures structure of our model clouds, but it is also the case that the agreement between \th\ and HINSA line widths means that there cannot be  significant saturation broadening taking place for the latter. 
This in turn implies that the HI density must be within a factor of a few of the steady state value in order that the peak HI absorption optical depths be $\leq$ unity.
In cases where the background antenna temperature is lower than the temperature in the outer part of the foreground cloud, excess emission above the background results.
In this situation, we see complex spectra, but it is always the case that a narrow absorption line traces the cold HI in the well--shielded cloud core within a cloud that has evolved to being fairly close to steady state.
Based on general comparison with observed spectral lines, an interval of a few $\times$ 10$^{14}$ s (1$\times$10$^7$ yr) can be set as the minimum that must have passed between the initial cloud compression and the present state of clouds as traced by HINSA.

The heating of gas in the cloud by the interstellar radiation field becomes significant in the outer portion of the cloud defined by $A_v$ $\leq$ 1 mag (see e.g. Figure 5 of Le Bourlot et al. 1993) for $\chi$ = 1 as appropriate to a cloud in a low--mass star forming region).  
It is not surprising that the sizes of regions defined by $T$ $\leq$ 12 K and $A_v$ $\leq$ 1 mag are quite similar.
The absolute size does depend on the cloud model, being smaller by a factor $\simeq$ 3.8 for Model 2 compared to Model 1, somewhat less than the factor 4.6 by which the central density is larger.
Clouds with different density profiles will not follow this relationship, but in the context of understanding the close agreement of clouds sizes measured in \th, \ce, and HINSA presented in detail in Paper II, it is important to realize that the ``molecular cloud'' size is effectively determined by the point where photochemistry of the trace species
becomes dominant.  
This again is where the external dust extinction is $\simeq$ 1 to a few mag.
We thus see a natural explanation of the clouds' having a similar size in cold HI and in carbon monoxide isotopologues.
The detailed agreement is affected by self shielding in \th\ and \ce, which along with chemical isotopic fractionation explains why the clouds sizes are larger in \th\ than \ce\ (Paper II).

The studies of HI narrow self absorption directly probe the central regions of dark clouds, but from the perspective of what the spectra do {\it not} look like, are relevant to the issue of the extent of lower density halos around molecular clouds, which may contain atomic (Andersson, Wannier, \& Morris 1991; Andersson, Roger, \& Wannier 1992) and molecular (Wannier et al. 1993) species.
Studies of these regions are difficult but as pointed out by \cite{bensch2006} these halos may have a significant effect on the interior regions of the clouds in terms of increasing the attenuation of the external radiation field as well as providing some pressure confinement of the molecular gas.

To interpret the observations of the cloud halos and cores in a unified manner will require combining realistic models of the cloud dynamics with accurate values for the rates of critical chemical reactions.  
Detailed investigation of spatial correlations between emission from cloud halos and absorption from cloud cores should be very valuable in assessing the evolutionary status of clouds.  
This requires at least two--dimensional models, including correction of the photodissociation rates for non--planar geometry as discussed by \cite{ford2003}.  
More detailed treatment of the HI $\rightarrow$ \h2\ conversion is also required, especially at late times when the density of atomic hydrogen is low (Biham et al. 2001; Biham \& Lipshtat 2002).
With this improved modeling, observations of HINSA in molecular clouds should become an effective tool for obtaining constraints on the evolutionary history of clouds, their age, and the overall time scale for star formation. 

This work was supported in part by the National Science Foundation through grant AST 0404770 to Cornell University, and by the Jet Propulsion Laboratory, California Institute of Technology. We thank the reviewer for a very careful reading of this manuscript and a large number of suggestions which improved its readability.  This research has made use of NASA's Astrophysics Data System.  



\end{document}